\documentclass{jfm}
\usepackage{graphicx}
\usepackage{newtxtext}
\usepackage{newtxmath}
\usepackage{natbib}
\usepackage{hyperref}
\usepackage{xcolor}
\usepackage{pifont}
\usepackage{tikz}
\usetikzlibrary{tikzmark}
\tikzset{mycircled/.style={circle,draw,inner sep=0.1em,line width=0.1em}}
\hypersetup{
    colorlinks = true,
    urlcolor   = blue,
    citecolor  = blue,
}

\newcommand{\RomanNumeralCaps}[1]
\linenumbers

\shorttitle{Journal of Fluid Mechanics}
\shortauthor{R. Lottem and E. Boyko}
\title{\Large Radial flow of an Oldroyd-B fluid: theoretical and simulation results in the ultra-dilute limit}
 \author{Ron Lottem\aff{1} \and 
 Evgeniy Boyko\aff{1}
 \corresp{\email\href{mailto:evgboyko@technion.ac.il}{evgboyko@technion.ac.il}}
}
  \affiliation{\aff{1}Faculty of Mechanical Engineering, Technion – Israel Institute of Technology,
Haifa 3200003, Israel}

\begin{document}
\maketitle

\begin{abstract}
Pressure-driven radial flows of viscoelastic fluids are common in various industrial applications, such as injection molding and extrusion. It is well known that viscoelastic rheology can significantly impact the hydrodynamic features of non-Newtonian flows. However, these hydrodynamic features remain not fully understood compared to those observed in Newtonian flows. We analyze the pressure-driven radial flow of an Oldroyd-B fluid between parallel plates and present a theoretical framework together with finite-element numerical simulations for determining the flow rate--pressure drop relation. Unlike previous theoretical studies restricted to the weakly viscoelastic limit of low Deborah ($De$) numbers, we apply lubrication theory and consider the ultra-dilute limit, which allows us to study viscoelastic radial flows at order-one Deborah numbers. Using the one-way coupling between the Newtonian velocity profile and elastic stresses, we derive closed-form expressions for the conformation tensor and pressure drop in the ultra-dilute limit. We show that the pressure drop monotonically increases with $De$, identify the physical mechanisms governing this increase, and delineate the range of validity of the ultra-dilute approximation. We further reveal that the low-$De$ asymptotic analysis of the pressure drop may fail to accurately capture finite-element simulation results, even at low Deborah numbers, owing to its inability to satisfy the inlet elastic stresses. In contrast, our theoretical predictions based on the ultra-dilute limit are in excellent agreement with the finite-element simulation results, enabling us to elucidate the pressure drop behavior for order‑one Deborah numbers.

 \end{abstract}


%{\bf MSC Codes }  {\it(Optional)} Please enter your MSC Codes here

\section{Introduction}

Pressure-driven radial flows of viscoelastic fluids arise in various polymer-processing applications, including injection molding and extrusion~\citep{middleman1977fundamentals,pearson,tadmor2013principles}. It is well known that even at low polymer concentrations, viscoelastic fluids can exhibit flow characteristics that differ significantly from those of Newtonian fluids. The stretching of polymer molecules by the flow generates elastic stresses in addition to viscous stresses, giving rise to nonlinear viscoelastic effects such as normal stress differences
and extensional thickening~\citep{bird1987dynamics1, steinberg2021elastic,datta2021perspectives,ewoldt2022designing}. Even at low Reynolds numbers, these viscoelastic effects may drastically change the hydrodynamic features of the flow, including the relationship between the pressure drop $\Delta p$ and the flow
rate $q$. Therefore, accurate prediction of the impact of fluid viscoelasticity on the hydrodynamic features of radial flows is important both for engineering applications and for advancing the fundamental understanding of non-Newtonian fluid mechanics.

Over the years, the effect of fluid viscoelasticity in radial-flow configurations has been studied through theoretical analyses~\citep{schwarz1969radial, lee_radial_1976, co1977slow} and experimental measurements~\citep{schwarz1969radial,laurencena1974radial,lee_radial_1976_exp}. For example, \citet{lee_radial_1976} and \citet{co1977slow} used the five-constant Oldroyd and third-order fluid models, respectively, to study radial flow in the weakly viscoelastic limit, corresponding to small Deborah ($De$) or Weissenberg ($Wi$) numbers (see~\S~\ref{Scaling} for definitions).
In addition to viscoelasticity, \citet{lee_radial_1976} and \citet{co1977slow}~considered the effects of weak fluid inertia and weak shear thinning, and predicted that both reduce the pressure drop in radial flow between parallel plates. By contrast, viscoelasticity has the opposite effect, increasing the pressure drop. Therefore, the net effect on pressure drop depends on the relative magnitudes of inertial, shear-thinning, and viscoelastic effects. To test the theoretical predictions of \citet{lee_radial_1976}, \cite{lee_radial_1976_exp} performed experiments in a radial-flow configuration and measured the radial pressure distribution of a polyacrylamide solution. The fluid was highly elastic but, unfortunately, also exhibited \emph{strong} shear-thinning behavior. As a result,~\cite{lee_radial_1976_exp} could not quantitatively compare their measurements with the available weakly viscoelastic theory, which accounted only for \emph{weak} shear-thinning effects.

Most theoretical studies of viscoelastic radial flows have applied lubrication theory to simplify the governing equations and facilitate the analysis \citep{schwarz1969radial,lee_radial_1976,co1977slow}. Such lubrication-based approaches are common in a wide range of viscoelastic flow problems, including pressure-driven flows in slowly varying channels~\citep{boyko2022pressure,housiadas2023lubrication, HinchBoykoStone2024, BoykoHinchStone2024,mahapatra2025viscoelastic,kedem2026viscoelastic}, viscoelastic thin-film boundary-driven lubrication and tribology~\citep{tichy1996non,sawyer1998non,ahmed2021new,gamaniel2021effect,ahmed2023modeling,sari2024effect, sari2025role,ahmed2025leveraging,AhmedBiancofiore2025}, free-surface flows~\citep{ro1995viscoelastic,zhang2002surfactant,saprykin2007free,datt2022thin}, and motion of a sphere near a rigid plane in a viscoelastic fluid~\citep{Ardekani_2007,Ruangkriengsin2024}.

Furthermore, many theoretical studies of viscoelastic flows have employed lubrication theory together with a perturbation expansion in powers of the Deborah number to obtain analytical results in the weakly viscoelastic limit. For example, \citet{boyko2022pressure} studied the steady flow of an Oldroyd-B fluid in a slowly varying contraction and derived asymptotic expressions for the dimensionless pressure drop up to $O(De^3)$ in the low-Deborah-number limit. \citet{housiadas2023lubrication} extended this analysis to substantially higher orders, obtaining asymptotic expressions for the pressure drop up to $O(De^8)$ for several constitutive models, including Oldroyd-B, Giesekus~\citep{giesekus1982simple}, Phan-Thien--Tanner (PTT)~\citep{thien1977new,phan1978nonlinear}, and the finitely extensible nonlinear elastic model with the Peterlin approximation (FENE-P)~\citep{bird1980polymer,bird1987dynamics1}.
Recently,~\citet{AhmedBiancofiore2025} studied the influence of fluid viscoelasticity on the hydrodynamics of thin-film lubrication contacts between finite-width sliding channels using the Oldroyd-B model in the low-Deborah-number limit.
In fact, the early studies of \citet{lee_radial_1976} and \citet{co1977slow} on the radial flow of viscoelastic fluids considered the weakly viscoelastic limit and derived asymptotic expressions for the pressure drop between parallel plates up to $O(De)$, predicting an increase in the pressure drop with increasing $De$ at low Deborah numbers.

Although low-Deborah-number asymptotic analyses often yield closed-form analytical expressions, their validity is restricted to weakly viscoelastic flows, and they cannot accurately capture flow behavior at order-one and high Deborah numbers, where elastic effects are significant. Another approach to simplifying the governing equations and analyzing viscoelastic flows at non-small Deborah numbers is to consider the ultra-dilute limit~\citep{remmelgas1999computational,koch2016stress,moore2012weak,li2019orientation,mokhtari2022birefringent,sharma2025extensional,BoykoHinchStone2024,HinchBoykoStone2024,mahapatra2025viscoelastic,hinch2026approach,kedem2026viscoelastic}, in which the polymer contribution to the viscosity is small, $\tilde{\beta}=\eta_p/\eta_0\ll1$, where $\eta_p$ is the polymer contribution to the total zero-shear-rate viscosity $\eta_0$ of the
polymer solution.
Physically, this limit corresponds to a low concentration of polymer molecules suspended in a Newtonian solvent~\citep{remmelgas1999computational,mokhtari2022birefringent}. In the ultra-dilute limit, the velocity field remains approximately Newtonian, resulting in a one-way coupling between the Newtonian velocity and the elastic stresses at the leading order in $\tilde{\beta}$. This one-way coupling greatly simplifies the theoretical analysis and enables the calculation of viscoelastic corrections to the velocity and pressure fields from the leading-order elastic stresses, thereby providing insight into the influence of fluid viscoelasticity even at order-one and high Deborah numbers. 

Previous studies employed the ultra-dilute limit to examine the structure of the stress field in flows past a cylinder~\citep{renardy2000asymptotic}, a sphere~\citep{moore2012weak}, and arrays of cylinders~\citep{mokhtari2022birefringent}, as well as in stagnation~\citep{becherer2009stress,van2009viscoelastic} and cross-slot~\citep{remmelgas1999computational} flows. The same framework has also been used to investigate the extensional rheology of dilute suspensions of spheres in viscoelastic fluids~\citep{sharma2025extensional} and to study the pressure drop of viscoelastic fluids in slowly varying contraction~\citep{BoykoHinchStone2024,HinchBoykoStone2024,mahapatra2025viscoelastic,hinch2026approach} and contraction--expansion~\citep{kedem2026viscoelastic} configurations. In particular, \citet{BoykoHinchStone2024}, \citet{HinchBoykoStone2024}, \citet{hinch2026approach} and \citet{kedem2026viscoelastic} recently combined the ultra-dilute limit with the lubrication approximation to analyze the flow of an Oldroyd-B fluid in slowly varying channels at order-one and high Deborah numbers. \citet{BoykoHinchStone2024} and \citet{kedem2026viscoelastic} derived closed-form semi-analytical expressions for the conformation tensor and pressure drop in slowly varying contraction and contraction--expansion channels for arbitrary values of the Deborah number. Furthermore, \citet{BoykoHinchStone2024}, \citet{HinchBoykoStone2024}, and \citet{hinch2026approach} obtained asymptotic solutions for the pressure drop in planar and axisymmetric contractions in the high-$De$ limit, showing that the pressure drop of an Oldroyd-B fluid monotonically decreases with $De$.

It should be noted that beyond viscoelastic fluids, numerous studies examined radial flows of shear-thinning and viscoplastic fluids~\citep[see, e.g.,][]{na_radial_1967,khader_inertia_1973,laurencena1974radial,co_inelastic_1981,dai_radial_1981,zou_radial_2020,shamu_radial_2020,albattat2021semi,ashkenazi2025radial}. For example, \citet{zou_radial_2020} analyzed the radial flow of a viscoplastic fluid between homogeneous fractures and derived an analytical solution for the flow based on the Herschel--Bulkley fluid model. Recently,~\citet{ashkenazi2025radial} studied the radial flow of a shear-thinning fluid between parallel plates and presented a theoretical framework for calculating the pressure distribution and pressure drop of an Ellis fluid model.
They compared their theoretical predictions with finite-element simulation results and the experimental data of \citet{laurencena1974radial}, obtaining excellent agreement in both cases. Moreover, the theory of~\citet{ashkenazi2025radial} accurately captures the interplay between the shear-thinning and zero-shear-rate effects on the pressure distribution and pressure drop, which cannot be described using a simple power-law model. 

In contrast to the modeling of radial flows of shear-thinning and viscoplastic fluids, which has enabled considerable theoretical progress over a wide range of flow rates \citep{zou_radial_2020,albattat2021semi,ashkenazi2025radial}, all theoretical studies of viscoelastic radial flows have thus far been restricted to small Deborah or Weissenberg numbers, corresponding to the weakly viscoelastic limit~\citep{lee_radial_1976,co1977slow}. Therefore, the behavior of viscoelastic radial flows at non-small
Deborah/Weissenberg numbers remains largely unexplored,
motivating further investigation.

In this work, we study the steady pressure-driven radial flow of an Oldroyd-B fluid between two parallel plates and present a theoretical framework, together with finite-element simulations, for predicting the flow rate--pressure drop relation.
In contrast to \citet{lee_radial_1976} and \cite{co1977slow}, who studied the viscoelastic radial flow in the low-Deborah-number limit, we focus on the ultra-dilute limit, which enables us to examine the velocity, elastic stresses, and pressure drop at order-one Deborah numbers. We apply the lubrication approximation and use a one-way coupling between the
Newtonian velocity profile and elastic stresses to derive semi-analytical expressions for the conformation
tensor and dimensionless pressure drop in
the ultra-dilute limit. These semi-analytical expressions allow us to study the variation of the elastic stresses and pressure drop behavior between the plates at order-one Deborah numbers. We provide an analytical expression for the pressure
drop of an Oldroyd-B fluid in the low-Deborah-number limit, which is consistent with previous results of~\citet{lee_radial_1976} and \cite{co1977slow}. Furthermore, we elucidate the physical mechanisms governing the pressure drop behavior and delineate the range of validity of the ultra-dilute approximation in radial-flow configurations. Our semi-analytical results in the ultra-dilute limit are in excellent agreement with finite-element simulations and advance our understanding of radial flows of viscoelastic fluids. We believe that these results are of fundamental importance because they can be directly compared with experimental measurements of constant shear-viscosity viscoelastic (Boger) fluids~\citep {james2009boger} in radial configurations.

\section{Problem formulation and governing equations}\label{PF}

\begin{figure}
\centerline{\includegraphics[width=\linewidth]{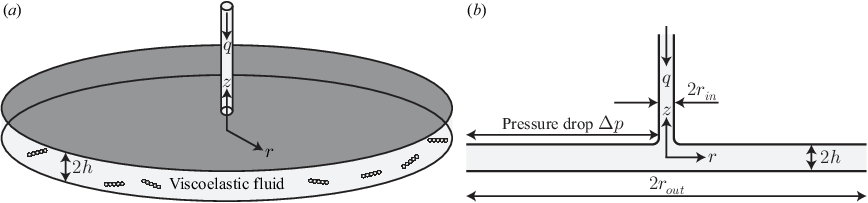}}
\caption{Schematic illustration of the flow configuration, showing the coordinate
system and relevant physical parameters. ($a$)  The radial flow of a viscoelastic fluid between two disk-shaped, parallel plates of radius $r_{\it out}$. Two plates are separated by a small gap $ 2 h$ ($h\ll r_{\it out}$) and contain a viscoelastic fluid steadily driven by the imposed flow rate $q$ through a narrow tube of radius $r_{\it in}$ at the center of the top plate. ($b$) Cross-section view of the geometry. Our interest is to determine the pressure drop $\Delta p$ between the inlet ($r=r_{\it in}$) and outlet ($r=r_{\it out}$).}\label{F1}
\end{figure}

We analyze the incompressible, steady, radial flow of a viscoelastic fluid between two disk-shaped plates of radius $r_{\it out}$ separated by a small gap $ 2 h$, where $h\ll r_{\it out}$.
We employ cylindrical coordinates $(r,z,\theta)$ and assume that the axisymmetric flow is driven by a constant volumetric flow rate $q$ through a narrow tube of radius $r_{\it in} \ll r_{\it out}$, which enters at the center of the top plate, as shown in figure~\ref{F1}. The imposed flow rate induces the radial fluid motion with velocity $\boldsymbol{u}=(u_{r},u_{z})$
and pressure distribution $p$.
Our primary interest is to determine the pressure drop $\Delta p$ between the inlet ($r=r_{\it in}$) and outlet ($r=r_{\it out}$) for a given $q$.

We consider low-Reynolds-number flows so that the fluid motion is governed by the continuity equation and Cauchy momentum equations in the absence of inertia
\refstepcounter{equation}
$$
\boldsymbol{\nabla\cdot u}=0,\qquad\boldsymbol{\nabla\cdot\sigma}=\boldsymbol{0},\eqno{(\theequation{a,b})}\label{Continuity+Momentum}
$$where $\boldsymbol{\sigma}$ is the stress tensor.

To describe the viscoelastic behavior of the fluid, we employ the Oldroyd-B constitutive model~\citep{oldroyd1950formulation}, which represents the simplest coupling of viscous and elastic stresses and is widely used to describe the flow of viscoelastic Boger fluids with constant shear viscosity. The Oldroyd-B equation can be derived from microscopic considerations by representing polymer molecules as elastic dumbbells that obey a linear Hooke’s law, undergoing advection and stretching by the flow. The corresponding stress tensor $\boldsymbol{\sigma}$ is 
\begin{equation}
\boldsymbol{\sigma}=-p\mathsfbi{I}+2\eta_s\mathsfbi{E}+\boldsymbol{\tau}_p,
\label{momentum}
\end{equation}
where the first term on the right-hand side of (\ref{momentum}) is the pressure contribution, the second term is the viscous stress contribution of a Newtonian solvent with a constant viscosity $\eta_s$, where $\mathsfbi{E}=\frac{1}{2}(\boldsymbol{\nabla{u}}+(\boldsymbol{\nabla{u}})^{ \rm T})$ is the rate-of-strain tensor, and the last term, $\boldsymbol{\tau}_p$, is the polymer contribution. 

For the Oldroyd-B model, the polymer contribution to the stress tensor $\boldsymbol{\tau}_p$ can be expressed in terms of the symmetric conformation tensor $\mathsfbi{A}$ as \citep{bird1987dynamics1,larson1988constitutive,Intro_C_F}
\begin{equation}
 \boldsymbol{\tau}_p= \frac{\eta_p}{\lambda}(\mathsfbi{A}-\mathsfbi{I}),
\label{tau_p}
\end{equation}
where $\lambda$ is the relaxation time and $\eta_p$ is the polymer contribution to the shear viscosity at zero shear rate. It is also convenient to introduce the total zero-shear-rate viscosity $\eta_0=\eta_s+\eta_p$.

At a steady state, the evolution equation for the conformation tensor $\mathsfbi{A}$ of the Oldroyd-B fluid is given as \citep{bird1987dynamics1,larson1988constitutive,Intro_C_F}
\begin{equation}
\boldsymbol{u}\boldsymbol{\cdot}\boldsymbol{\nabla}\mathsfbi{A}-(\boldsymbol{\nabla}\boldsymbol{u})^{\mathrm{T}}\boldsymbol{\cdot}\mathsfbi{A}-\mathsfbi{A}\boldsymbol{\cdot}(\boldsymbol{\nabla}\boldsymbol{u})=-\frac{1}{\lambda}(\mathsfbi{A}-\mathsfbi{I}).
\label{A}
\end{equation}
For the fluid flow between the plates, the integral constraint for the flow rate is given as~\citep{ashkenazi2025radial}
\begin{equation}
2\pi\int_{-h}^{h} u_{r} r \mathrm{d}z=q.
\label{mass_cons}
\end{equation}
From (\ref{mass_cons}), it follows that the radial velocity has the form of $ u_{r}(r,z)=g(z)/r$~\citep[see, e.g.][]{park_flow_2020}. Substituting this result into the continuity equation (\ref{Continuity+Momentum}$a$) and using the no-penetration boundary conditions implies that $u_z\equiv0$, so that the flow field between the two disk-shaped plates is $\boldsymbol{u}=u_{r}(r,z) \boldsymbol{e}_r$.

\subsection{Non-dimensionalization}\label{Scaling}

We consider a narrow configuration, in which $h\ll r_{\it out}$, where $h$ is the half-height between the disks. Therefore, for the non-dimensionalization, we introduce the non-dimensional variables based on lubrication theory \citep{tichy1996non,zhang2002surfactant,ahmed2021new,boyko2022pressure,Ruangkriengsin2024,boyko2025pressureFD,boyko2025interplay,AhmedBiancofiore2025},
\begin{subequations}
\begin{equation}
R=\frac{r}{r_{\it out}}, \quad Z=\frac{z}{h},\quad U_{r}=\frac{u_{r}}{u_{c}}, \quad P=\frac{p}{\eta_0 u_cr_{\it out}/h^2} \quad \Delta P=\frac{\Delta p}{\eta_0 u_cr_{\it out}/h^2},
\label{ND_var1}
\end{equation}
\begin{equation}
 \mathcal{T}_{\it p,rr}=\frac{\tau_{\it p,rr}}{\eta_0u_cr_{\it out}/h^2},
\quad
 \mathcal{T}_{\it p,rz}=\frac{\tau_{\it p,rz}}{\eta_0u_c/h},
 \quad
 \mathcal{T}_{\it p,zz}=\frac{\tau_{\it p,zz}}{\eta_0 u_c/r_{\it out}},
 \quad 
\mathcal{T}_{\it p,\theta\theta}=\frac{\tau_{\it p,\theta\theta}}{\eta_0u_cr_{\it out}/h^2},
\label{ND_var2}
\end{equation}
\begin{equation} 
\mathcal{A}_{\it rr}=\epsilon^2A_{\it rr},
\quad
\mathcal{A}_{\it rz}=\epsilon A_{\it rz},
\quad
\mathcal{A}_{\it zz}=A_{\it zz}, 
\quad
\mathcal{A}_{\it \theta\theta}=\epsilon^2A_{\it \theta\theta},
\label{ND_var3}
\end{equation}\label{ND_variables}\end{subequations}
where $u_{c}={q/(2\pi h r_{\it out})}$ is  the characteristic velocity scale.

In addition, we introduce the aspect ratio of the
configuration, $\epsilon$, and the inlet-to-outlet aspect ratio, $\alpha$, which are assumed to be small,
\begin{equation}
 \epsilon=\frac{h}{r_{\it out}}\ll1 \quad \text{and} \quad \alpha=\frac{r_{\it in}}{r_{\it out}}\ll1, 
\label{Epsilon and alpha}
\end{equation}
the viscosity ratios,
\begin{equation}
\tilde{\beta}=\frac{\eta_p}{\eta_p+\eta_s}=\frac{\eta_p}{\eta_0} \quad \text{and} \quad \beta=1-\tilde{\beta}=\frac{\eta_s}{\eta_0},
\label{beta and beta_t}
\end{equation}
and the Deborah and Weissenberg numbers
\begin{equation}
De=\frac{\lambda u_c}{r_{\it out}} \quad \text{and} \quad Wi=\frac{\lambda u_c}{h}.
\label{De and Wi}
\end{equation}
The Deborah number $De$ is the ratio of the relaxation time of the fluid, $\lambda$, to the residence time between the plates, $r_{\it out}/u_c$~\citep{ahmed2021new,boyko2021RT,boyko2022pressure,housiadas2023lubrication}. In addition, we can introduce the Weissenberg number $Wi$ as the product of the relaxation time scale of the fluid, $\lambda$, and the characteristic shear rate of the flow, $u_c/h$. Therefore, the Deborah and Weissenberg numbers are related through $De = \epsilon Wi$, and for viscoelastic lubrication flows in narrow configurations with $\epsilon \ll 1$, $De$ can be small, but $Wi = O(1)$. 

\subsection{Non-dimensional governing equations in the lubrication limit}

Substituting the non-dimensional variables (\ref{ND_variables})--(\ref{De and Wi}) into the governing equations (\ref{Continuity+Momentum})--(\ref{A}), using the assumption that $\boldsymbol{u}=u_{r}\boldsymbol{e}_r$, and considering the leading order in $\epsilon$, we obtain (see Appendix~\ref{App A})
\begin{subequations}
\begin{equation}
\frac{\partial P}{\partial R}= (1-\tilde{\beta})\frac{\partial^2 U_{\it r}}{\partial Z^2}+\frac{\tilde{\beta}}{De} \left( \frac{1}{R}\frac{\partial(R\mathcal{A}_{\it rr)}}{\partial R}+\frac{\partial \mathcal{A}_{\it rz}}{\partial Z}-\frac{\mathcal{A}_{\theta\theta}}{R}\right),
\label{DPDR}
\end{equation}
\begin{equation}
\frac{\partial P}{\partial Z}=0,
\label{DPDZ}
\end{equation} 
\begin{equation}
U_{r} \frac{\partial \mathcal{A}_{\it zz}}{\partial R}= -\frac{1}{De}( \mathcal{A}_{\it zz}-1),
\label{Azz}
\end{equation}
\begin{equation}
U_{r}\frac{\partial \mathcal{A}_{\theta\theta}}{\partial R}-2\frac{U_{r}}{R}\mathcal{A}_{\theta\theta}=-\frac{1}{De}\mathcal{A}_{\theta\theta},\label{Athth} 
\end{equation}
\begin{equation}
U_{r} \frac{\partial \mathcal{A}_{\it rz}}{\partial R}-\frac{\partial U_{r}}{\partial Z} \mathcal{A}_{\it zz}+\frac{U_{r}}{R}\mathcal{A}_{\it rz} = -\frac{1}{De}\mathcal{A}_{\it rz},
\label{Arz}
\end{equation}
\begin{equation}
U_{r} \frac{\partial \mathcal{A}_{\it rr}}{\partial R}-2\frac{\partial U_{r}}{\partial Z} \mathcal{A}_{\it rz}-2\frac{\partial U_{r}}{\partial R} \mathcal{A}_{\it rr} = -\frac{1}{De}\mathcal{A}_{\it rr}.
\label{Arr}
\end{equation}\label{ND_gov}\end{subequations}From (\ref{DPDZ}), it follows that $P = P(R)$, i.e., the pressure is independent of $Z$ up to $O(\epsilon^2)$, consistent with the classical lubrication approximation. As shown in Appendix~\ref{App A}, the scaled $\mathcal{A}_{\theta\theta}$ and $\mathcal{A}_{\it rr}$ on the right-hand side of (\ref{Athth}) and (\ref{Arr}), respectively,  relax to $\epsilon^2$, which is neglected at the leading order in $\epsilon$.

We note that that the expressions for $ \mathcal{T}_{\it p,zz}$, $ \mathcal{T}_{\it p,\theta \theta}$, $\mathcal{T}_{\it p,rz}$, and $\mathcal{T}_{\it p,rr}$  are related to the corresponding conformation tensor components $\mathcal{A}_{\it zz}$, $\mathcal{A}_{\it \theta\theta}$, $\mathcal{A}_{\it rz}$, and $\mathcal{A}_{\it rr}$ through
\begin{subequations}\begin{equation}
\mathcal{T}_{\it p,zz}=\frac{\tilde{\beta}}{De}(\mathcal{A}_{\it zz}-1),\qquad
\mathcal{T}_{\it p,\theta\theta}=\frac{\tilde{\beta}}{De}\mathcal{A}_{\it \theta\theta}+O(\epsilon^2),
\end{equation}
\begin{equation}
\mathcal{T}_{\it p,rz}=\frac{\tilde{\beta}}{De}\mathcal{A}_{\it rz}, \qquad
\mathcal{T}_{\it p,rr}=\frac{\tilde{\beta}}{De}\mathcal{A}_{\it rr}+O(\epsilon^2).
\end{equation}\label{Polymer stresses via A}\end{subequations}
The corresponding boundary conditions on the velocity are
\refstepcounter{equation}
$$
U_{r}(R,Z=\pm1)=0, \qquad \int_{-1}^{1} U_{r} R\space \mathrm{d}Z=1,\eqno{(\theequation{a,b})}\label{BC ND velocity}
$$which represent, respectively, the no-slip boundary conditions at the plates and the
integral constraint for the flow rate.

The Oldroyd-B constitutive equations (\ref{Azz})--(\ref{Arr}) are hyperbolic and thus require knowledge of the values of the conformation tensor at the inlet, $R=\alpha$. 
For pressure-driven flow of the Oldroyd-B fluid in slowly varying contraction or contraction--expansion geometries, previous studies have prescribed inlet values of the conformation tensor corresponding to the fully developed Poiseuille flow in a straight channel upstream of the contraction~\citep[see, e.g.,][]{boyko2022pressure,BoykoHinchStone2024,HinchBoykoStone2024,mahapatra2025viscoelastic,hinch2026approach,kedem2026viscoelastic}. For radial flow, however, determining the inlet conformation tensor distribution at $R = \alpha$ is not straightforward, making the analytical treatment much more complicated.
In this study, we assume that we know these inlet reference distributions, which are obtained from finite-element simulations,
\begin{subequations}
\begin{equation}
\mathcal{A}_{\it zz}(R=\alpha,Z)=\mathcal{A}_{\mathit{zz}}^{\mathit{ref}}(Z), \qquad \mathcal{A}_{\it \theta \theta}(R=\alpha,Z)=\mathcal{A}_{\theta \theta}^{\mathit{ref}}(Z),
\end{equation}
\begin{equation}
\mathcal{A}_{\it rz}(R=\alpha,Z)=\mathcal{A}_{\mathit{rz}}^{\mathit{ref}}(Z), \qquad \mathcal{A}_{\it rr}(R=\alpha,Z)=\mathcal{A}_{\it rr}^{\mathit{ref}}(Z),
\end{equation}
\label{BCs inlet}\end{subequations}
Finally, we note that, in general,  $\mathcal{A}_{\mathit{zz}}^{\mathit{ref}}(Z)$, $\mathcal{A}_{\mathit{\theta \theta}}^{\mathit{ref}}(Z)$, $\mathcal{A}_{\mathit{rz}}^{\mathit{ref}}(Z)$, and $\mathcal{A}_{\mathit{rr}}^{\mathit{ref}}(Z)$  may depend on the Deborah number and exhibit different dependencies on $De$; see discussion in~$\mathsection$~\ref{Results}. 

\subsection{Non-dimensional pressure drop in the lubrication limit}

In this subsection, we calculate the pressure drop without solving directly for the velocity field. We integrate by parts the integral constraint  (\ref{BC ND velocity}$b$), repeatedly, using (\ref{BC ND velocity}$a$),
\begin{equation}
\frac{1}{R}=\int_{-1}^{1}U_{r}\mathrm{d}Z=\underset{0}{\underbrace{\left.\frac{1}{2}(1-Z^{2})\frac{\partial U_{r}}{\partial Z}\right|_{-1}^{1}}}-\frac{1}{2}\int_{-1}^{1}(1-Z^{2})\frac{\partial^{2}U_{r}}{\partial Z^{2}}\mathrm{d}Z.\label{Integration by parts, repeatedly_gen}
\end{equation}
Substituting the expression for $\partial^{2}U_{r}/\partial Z^{2}$ from (\ref{DPDR}) into (\ref{Integration by parts, repeatedly_gen}), we obtain 
\begin{equation}
 -\frac{1-\tilde{\beta}}{R}=\frac{1}{2}\int_{-1}^{1}(1-Z^{2})\left[\frac{\mathrm{d}P}{\mathrm{d}R}-\frac{\tilde{\beta}}{De}\left(\frac{1}{R}\frac{\partial(R\mathcal{A}_{\it rr})}{\partial R}+\frac{\partial\mathcal{A}_{\it rz}}{\partial Z}-\frac{\mathcal{A}_{\theta\theta}}{R}\right)\right]\mathrm{d}Z.\label{dP/dZ_gen}
\end{equation}
Recalling that $P=P(R)$, (\ref{dP/dZ_gen}) can be rearranged to yield the expression for the pressure gradient
\begin{equation}
    \frac{\mathrm{d}P}{\mathrm{d}R}=-\frac{3(1-\tilde{\beta})}{2R}+\frac{3 \tilde{\beta}}{4De}\int_{-1}^{1}(1-Z^{2})\left[\frac{1}{R}\frac{\partial(R\mathcal{A}_{\it rr})}{\partial R}+\frac{\partial\mathcal{A}_{\it rz}}{\partial Z}-\frac{\mathcal{A}_{\theta\theta}}{R}\right]\mathrm{d}Z. \label{dP/dR expl_gen}
\end{equation}
Integrating (\ref{dP/dR expl_gen}) with respect to $R$ from $\alpha$ to $1$ provides the pressure drop $\Delta  P\equiv P(\alpha)-P(1)$ between the plates 
\begin{eqnarray}
     \Delta P&=&-\int_{\alpha}^{1}\frac{\mathrm{d}P}{\mathrm{d}R}\mathrm{d}R=\frac{3}{2}(1-\tilde{\beta})\ln{\left(\frac{1}{\alpha}\right)}
-\frac{3 \tilde{\beta}}{4De} \int_{\alpha}^{1}\left[\int_{-1}^{1}(1-Z^{2})\frac{\partial\mathcal{A}_{\it rz}}{\partial Z}\mathrm{d}Z\right]\mathrm{d}R
\nonumber \\
&&-\frac{3 \tilde{\beta}}{4De}\int_{\alpha}^{1}\left[\int_{-1}^{1}(1-Z^{2})\left(\frac{1}{R}\frac{\partial(R\mathcal{A}_{\it rr})}{\partial R}-\frac{\mathcal{A}_{\theta\theta}}{R}\right)\mathrm{d}Z\right]\mathrm{d}R.\label{ND pressure drop 1_gen}
\end{eqnarray}
Using integration by parts, the non-dimensional pressure drop (\ref{ND pressure drop 1_gen}) can be expressed as 
\begin{eqnarray}
      \Delta P&=&\underset{\text{Solvent stress}}{\underbrace{\frac{3}{2}(1-\tilde{\beta})\ln{\left(\frac{1}{\alpha}\right)}}}
+\underset{\text{Elastic shear stress}}{\underbrace{-\frac{3 \tilde{\beta}}{2De} \int_{\alpha}^{1}\left[ \int_{-1}^{1}Z\mathcal{A}_{\it rz}\mathrm{d}Z \right]\mathrm{d}R}}
\nonumber \\
&&\underset{\text{Elastic normal stress}}{\underbrace{-\frac{3 \tilde{\beta}}{4De} \int_{-1}^{1}\left[(1-Z^{2})\left( \left.\mathcal{A}_{\it rr}\right|_{R=\alpha}^{R=1}+\int_{\alpha}^{1}\frac{\mathcal{A}_{\it rr}-\mathcal{A}_{\theta\theta}}{R}  \mathrm{d}R\right)\right]\mathrm{d}Z}},\label{ND pressure_drop_gen} 
\end{eqnarray}
where we have used the definition $\left.\mathcal{A}_{\it rr}\right|_{R=\alpha}^{R=1}=\mathcal{A}_{\it rr}(R=1,Z)-\mathcal{A}_{\it rr}(R=\alpha,Z)$.

Equation (\ref{ND pressure_drop_gen}) indicates that the non-dimensional pressure drop consists of
three contributions.
The first term on the right-hand side of (\ref{ND pressure_drop_gen}) represents the viscous contribution of the Newtonian solvent to the pressure drop. The second term represents the contribution of the elastic shear stresses, and the last term represents the contribution of the elastic normal stresses to the pressure drop.

\section{Low-Deborah-number lubrication analysis}\label{Low-Deborah-number lubrication analysis}

In $\mathsection$~\ref{PF}, we obtained the dimensionless equations (\ref{ND_gov}), which are governed by three non-dimensional parameters $\alpha$, $\tilde{\beta}$, and $De$, in the lubrication limit ($\epsilon \ll 1$). In this section, we derive analytical expressions for the velocity, conformation tensor, and the pressure drop for the radial flow of a weakly viscoelastic Oldroyd-B fluid at low Deborah numbers, $De \ll1$. To this end, we seek solutions of the form
\begin{equation}
\begin{pmatrix} U_{r} \\P\\\Delta P \\ \mathcal{A}_{\it zz}\\ \mathcal{A}_{\theta\theta} \\ \mathcal{A}_{\it rz} \\ \mathcal{A}_{\it rr} \end{pmatrix}=
\begin{pmatrix} U_{r}^{(0)} \\P^{(0)}\\\Delta P^{(0)} \\ \mathcal{A}_{\it zz}^{(0)}\\ \mathcal{A}_{\theta\theta}^{(0)}\\ \mathcal{A}_{\it rz}^{(0)} \\ \mathcal{A}_{\it rr}^{(0)} \end{pmatrix}+
De\begin{pmatrix} U_{r}^{(1)} \\P^{(1)}\\\Delta P^{(1)} \\ \mathcal{A}_{\it zz}^{(1)}\\ \mathcal{A}_{\theta\theta}^{(1)}\\ \mathcal{A}_{\it rz}^{(1)} \\ \mathcal{A}_{\it rr}^{(1)}
\end{pmatrix}
+O(De^2),
\label{Low De asymptotic}
\end{equation}
and derive asymptotic expressions for the pressure drop up to $O(De)$. 

At the leading order in $De$, the velocity profile is Newtonian 
\begin{equation}
U_{r}^{(0)}=\frac{3}{4}\frac{1}{R}(1-Z^2).
\label{U0_nonsym}
\end{equation} 
The corresponding conformation tensor components at the leading and first order in $De$ are
\begin{equation}
\mathcal{A}_{\it zz}^{(0)}=1, \qquad \mathcal{A}_{\it \theta\theta}^{(0)}=0, \qquad
\mathcal{A}_{\it rz}^{(0)}=0, \qquad
\mathcal{A}_{\it rr}^{(0)}=0,
\label{lowDe_A_0}
\end{equation}  
\begin{equation}
\mathcal{A}_{\it zz}^{(1)}=0, \qquad \mathcal{A}_{\it \theta\theta}^{(1)}=0, \qquad
\mathcal{A}_{\it rz}^{(1)}=\frac{\partial U_{r}^{(0)}}{\partial Z}=-\frac{3Z}{2R}, \qquad
\mathcal{A}_{\it rr}^{(1)}=0.
\label{lowDe_A_1_nonsym}
\end{equation}
Substituting (\ref{lowDe_A_1_nonsym}) into (\ref{ND pressure_drop_gen}), we obtain the leading-order Newtonian pressure drop 
\begin{equation}
\Delta P^{(0)}=  \frac{3}{2}(1-\tilde{\beta})\ln{\left(\frac{1}{\alpha}\right)}
-\frac{3}{2} \tilde{\beta} \int_{\alpha}^{1}\left[ \int_{0}^{1}Z\mathcal{A}_{\it rz}^{(1)}\mathrm{d}Z \right]\mathrm{d}R=\frac{3}{2}\ln{\left(\frac{1}{\alpha}\right)}.
\label{dP0_nonsym}
\end{equation}
To calculate the pressure drop at the first order in $De$ requires knowing $\mathcal{A}_{\it \theta\theta}^{(2)}$, $\mathcal{A}_{\it rr}^{(2)}$, and $\mathcal{A}_{\it rz}^{(2)}$. Substituting (\ref{Low De asymptotic}) into (\ref{Athth})--(\ref{Arr}) and considering the second order in $De$, we obtain
\begin{subequations}
\begin{equation}
\mathcal{A}_{\it \theta\theta}^{(2)}= 0,
\label{Athth2}
\end{equation}
\begin{equation}
\mathcal{A}_{\it rr}^{(2)}= 2\left(\frac{\partial U_{r}^{(0)}}{\partial Z}\right)^2= \frac{9Z^2}{2R^2},
\label{Arr2_nonsym}
\end{equation}
\begin{equation}
\mathcal{A}_{\it rz}^{(2)}= \frac{\partial U_{r}^{(1)}}{\partial Z}- \underset{0}{\underbrace{U_{r}^{(0)}\frac{\partial^2 U_{r}^{(0)}}{\partial Z\partial R}-\frac{U_{r}^{(0)}}{R} \frac{\partial U_{r}^{(0)}}{\partial Z}}}=\frac{\partial U_{r}^{(1)}}{\partial Z}.\label{Arz2}
\end{equation}
\label{lowDe_A_2}\end{subequations}
Substituting (\ref{Low De asymptotic}) into (\ref{ND pressure_drop_gen}) and using (\ref{Athth2}), the pressure drop at the first order in $De$ can be expressed as 
\begin{equation}
\Delta P^{(1)}= -\frac{3 \tilde{\beta}}{4} \int_{-1}^{1}\left[(1-Z^{2})\left( \left.\mathcal{A}_{\it rr}^{(2)}\right|_{R=\alpha}^{R=1}+\int_{\alpha}^{1}\frac{\mathcal{A}_{\it rr}^{(2)}}{R}  \mathrm{d}R\right)\right]\mathrm{d}Z -\frac{3}{2} \tilde{\beta} \int_{\alpha}^{1}\left[ \int_{-1}^{1}Z\mathcal{A}_{\it rz}^{(2)}\mathrm{d}Z \right]\mathrm{d}R.
\label{pressure_drop at O(De)_nonsym}
\end{equation}
The first term on the right-hand side of (\ref{pressure_drop at O(De)_nonsym}) is the contribution of the elastic normal stresses and can be easily calculated. In contrast, the second term, which is the contribution of elastic shear stresses, contains $U_{r}^{(1)}$ that is unknown. However, using the integral constraint at $O(De)$ and integration by parts, we obtain 
\begin{equation}
\int_{-1}^{1}U_{r}^{(1)}R \mathrm{d}Z=0=-\int_{-1}^{1}Z \frac{\partial U_{r}^{(1)}}{\partial{Z}} \mathrm{d}Z=-\int_{-1}^{1}Z \mathcal{A}_{\it rz}^{(2)}\mathrm{d}Z,
\end{equation} and thus, the last term in (\ref{pressure_drop at O(De)_nonsym}) vanishes. 
Therefore, using (\ref{Arr2_nonsym}) and (\ref{pressure_drop at O(De)_nonsym}), the pressure drop at $O(De)$ is
\begin{equation}
\Delta P^{(1)}= \frac{9}{20}\tilde{\beta}\left( \frac{1}{\alpha^2}-1\right) \quad \text{for} \quad De\ll1,
\label{PD_LowDe_nonsym}
\end{equation}
so that the total pressure drop in the low-$De$ limit, accounting for the leading-order effect of viscoelasticity, is
\begin{equation}
\Delta P= \frac{3}{2}\ln{\left(\frac{1}{\alpha}\right)}+ \frac{9}{20}De\tilde{\beta}\left( \frac{1}{\alpha^2}-1\right) \quad \text{for} \quad De\ll1.
\label{total_PD-low De_nonsym}
\end{equation}
Equation (\ref{total_PD-low De_nonsym}) clearly shows that, at low Deborah numbers, the dimensionless pressure drop monotonically increases with $De$, consistent with previous studies on the radial flow of viscoelastic fluids~\citep{lee_radial_1976,co1977slow}. 
Specifically, our low-$De$ expression for the pressure drop of the Oldroyd-B fluid (\ref{total_PD-low De_nonsym}) is in agreement with the expressions (33) and (35) of \citet{lee_radial_1976} and expression (40) of \citet{co1977slow} when accounting for the differences in non-dimensionalization. \citet{lee_radial_1976} and \citet{co1977slow} employed the five-constant Oldroyd model and the third-order fluid model, respectively, to analyze weakly viscoelastic radial flows while incorporating the effects of weak fluid inertia and shear thinning.

\section{Lubrication analysis in the ultra-dilute limit}\label{Lubrication analysis in the ultra-dilute limit}

In the previous section, we derived an analytical expression for the non-dimensional pressure drop of an Oldroyd-B fluid in a radial flow in the low-Deborah-number limit, similar to earlier studies of weakly viscoelastic radial flows with $De\ll1$~\citep{schwarz1969radial, lee_radial_1976,co1977slow}. However, the low-Deborah-number asymptotic analysis cannot accurately capture the viscoelastic behavior
at non-small Deborah numbers where there are significant elastic stresses~\citep[see, e.g.,][]{BoykoHinchStone2024,HinchBoykoStone2024,kedem2026viscoelastic}. 

To understand the viscoelastic behavior at non-small Deborah numbers, we assume $De = O(1)$ and consider the ultra-dilute limit, $\tilde{\beta}=\eta_p/\eta_0\ll1$~\citep{remmelgas1999computational,moore2012weak,li2019orientation,mokhtari2022birefringent,BoykoHinchStone2024,HinchBoykoStone2024,kedem2026viscoelastic}. In the ultra‑dilute limit, the problem exhibits a one‑way coupling between the velocity and pressure fields and the elastic stresses, represented by the conformation tensor. At leading order in $\tilde{\beta}$, the velocity and pressure remain Newtonian and unaffected by elastic stresses. The elastic stresses, manifested through the spatial evolution of the conformation tensor, are generated by this velocity field. As we show in the next
subsections, this one‑way coupling allows us to derive closed‑form asymptotic expressions for the conformation tensor at $O(\tilde{\beta}^{0})$ and the pressure drop up to $O(\tilde{\beta})$. To this end, we seek solutions of the form
\begin{equation}
\begin{pmatrix} U_{r} \\P\\\Delta P \\ \mathcal{A}_{\it zz}\\ \mathcal{A}_{\theta\theta} \\ \mathcal{A}_{\it rz} \\ \mathcal{A}_{\it rr} \end{pmatrix}=
\begin{pmatrix} U_{r,0} \\P_0\\\Delta P_0 \\ \mathcal{A}_{\mathit{\it zz},0}\\ \mathcal{A}_{\theta\theta,0}\\ \mathcal{A}_{\mathit{\it rz},0} \\ \mathcal{A}_{\mathit{\it rr},0} \end{pmatrix}+
\tilde{\beta}\begin{pmatrix} U_{r,1} \\P_1\\\Delta P_1 \\ \mathcal{A}_{\mathit{\it zz},1}\\ \mathcal{A}_{\theta\theta,1} \\ \mathcal{A}_{\mathit{\it rz},1} \\ \mathcal{A}_{\mathit{\it rr},1} 
\end{pmatrix}
+O(\tilde{\beta}^2,\epsilon^2).
\label{Asympt_beta_t}
\end{equation}

\subsection{Velocity, conformation, and pressure drop at the leading order in \texorpdfstring{$\tilde{\beta}$}{}}

Substituting (\ref{Asympt_beta_t}) into (\ref{DPDR}) and considering the leading order in $\tilde{\beta}$, the radial momentum equation takes the from
\begin{equation}
    \frac{\mathrm{d}P_0}{\mathrm{d}R}=\frac{\partial^2 U_{r,0}}{\partial Z^2},
    \label{DP0DR}
\end{equation}
subject to the boundary conditions (\ref{BC ND velocity}). 
The radial velocity $U_{r,0}$ and the pressure drop $\Delta P_0$ at the leading order in $\tilde{\beta}$ correspond to a Newtonian fluid flow and are given by 
\refstepcounter{equation}
$$
U_{r,0}=\frac{3}{4}\frac{1}{R}(1-Z^2)\qquad\hbox{and}\qquad \Delta P_0 =\frac{3}{2}\ln{\left(\frac{1}{\alpha}\right)},\eqno{(\theequation{a,b})}\label{U0 and dP0 ND}
$$consistent  with (\ref{U0_nonsym}) and (\ref{dP0_nonsym}).

Substituting (\ref{Asympt_beta_t}) and (\ref{U0 and dP0 ND}$a$) into (\ref{Athth})--(\ref{Arr}), and considering the leading order in $\tilde{\beta}$, the equations for the conformation tensor components simplify to
\begin{subequations}
\begin{equation}
U_{r,0}\frac{\partial \mathcal{A}_{\theta\theta,0}}{\partial R}-2\frac{U_{r,0}}{R}\mathcal{A}_{\theta\theta,0}=-\frac{1}{De}\mathcal{A}_{\theta\theta,0},\label{Athth_out} 
\end{equation}
\begin{equation}
U_{r,0} \frac{\partial \mathcal{A}_{\mathit{zz},0}}{\partial R}= -\frac{1}{De}( \mathcal{A}_{\mathit{zz},0}-1),
\label{Azz_out}
\end{equation}
\begin{equation}
U_{r,0} \frac{\partial \mathcal{A}_{\mathit{rz},0}}{\partial R}-\frac{\partial U_{r,0}}{\partial Z} \mathcal{A}_{\mathit{zz},0}+\frac{U_{r,0}}{R}\mathcal{A}_{\mathit{rz},0} = -\frac{1}{De}\mathcal{A}_{\mathit{rz},0},
\label{Arz_out}
\end{equation}
\begin{equation}
U_{r,0} \frac{\partial \mathcal{A}_{\mathit{rr},0}}{\partial R}-2\frac{\partial U_{r,0}}{\partial Z} \mathcal{A}_{\mathit{rz},0}-2\frac{\partial U_{r,0}}{\partial R} \mathcal{A}_{\mathit{rr},0} = -\frac{1}{De}\mathcal{A}_{\mathit{rr},0}.
\label{Arr_out}
\end{equation}\label{ND_gov_low_bt_out}\end{subequations}
subject to the boundary conditions
\begin{subequations}
\begin{equation}
\mathcal{A}_{\mathit{\theta\theta},0}(R=\alpha,Z)=\mathcal{A}_{\mathit{\theta\theta}}^{\mathit{ref}}(Z),
\label{bc_thth_out}
\end{equation}
\begin{equation}
\mathcal{A}_{\mathit{zz},0}(R=\alpha,Z)=\mathcal{A}_{\mathit{zz}}^{\mathit{ref}}(Z),
\label{bc_zz_out}
\end{equation}
\begin{equation}
\mathcal{A}_{\mathit{rz},0}(R=\alpha,Z)=\mathcal{A}_{\mathit{rz}}^{\mathit{ref}}(Z),
\label{bc_rz_out}
\end{equation}
\begin{equation}
\mathcal{A}_{\mathit{rr},0}(R=\alpha
,Z)=\mathcal{A}_{\mathit{rr}}^{\mathit{ref}}(Z).
\label{bc_rr_out}
\end{equation}\label{bc_b_out}\end{subequations}We note that, in the ultra‑dilute limit, the equation governing the $\theta \theta$-component of the conformation tensor, (\ref{Athth_out}), is decoupled from the remaining equations, whereas (\ref{Azz_out})--(\ref{Arr_out}) form a set of one-way coupled partial
differential equations. Thus, the equations for the conformation tensor components (\ref{Athth_out})--(\ref{Arr_out}) may be solved sequentially, first for $\mathcal{A}_{\mathit{\theta\theta},0}$ and $\mathcal{A}_{\mathit{zz},0}$, then for $\mathcal{A}_{\mathit{rz},0}$, and finally for $\mathcal{A}_{\mathit{rr},0}$.

Solving (\ref{ND_gov_low_bt_out}) together with the boundary conditions (\ref{bc_b_out}), we obtain closed-form expressions
for $\mathcal{A}_{\mathit{\theta\theta},0}$, $\mathcal{A}_{\mathit{zz},0}$, $\mathcal{A}_{\mathit{rz},0}$, and $\mathcal{A}_{\mathit{rr},0}$ for arbitrary values of $De$ 
\begin{subequations}
        \begin{equation}
    \frac{\mathcal{A}_{\theta\theta,0}}{R^2/\alpha^2}=\mathcal{A}_{\theta\theta}^{\it ref}(Z) \exp \left(\frac{f(R,Z)}{De}\right),
      \label{norm_Athth0b_out_exp}
    \end{equation}
    \begin{equation}
      \mathcal{A}_{\mathit{zz},0}=1+ (\mathcal{A}_{\it zz}^{\it ref}(Z)-1) \exp \left(\frac{f(R,Z)}{De}\right),
      \label{norm_Azz0b_out_exp}
    \end{equation}
        \begin{equation}
      \frac{\mathcal{A}_{\mathit{rz},0}}{3DeZ/2R}=-1+ \exp \left(\frac{f(R,Z)}{De}\right) \left(\frac{ \mathcal{A}_{\it rz}^{\it ref}(Z)}{3DeZ/2\alpha}
      +\frac{f(R,Z)}{De}(\mathcal{A}_{\it zz}^{\it ref}(Z)-1)
      +1\right),
      \label{norm_Arz0b_out_exp}
    \end{equation}
        \begin{eqnarray}
     \frac{\mathcal{A}_{\mathit{rr},0}}{9De^2Z^2/2R^2}&=& 1
   +\exp \left(\frac{f(R,Z)}{De}\right) \left[\frac{\mathcal{A}_{\it rr}^{\it ref}(Z)}{9De^2Z^2/2\alpha^2}\right. \nonumber \\ 
  &&\left. +\frac{f(R,Z)}{De}\left(\frac{\mathcal{A}_{\it rz}^{\it ref}(Z)}{3DeZ/2\alpha} 
     +\frac{f(R,Z)}{2De}(\mathcal{A}_{\it zz}^{\it ref}(Z)-1)+1\right)-1\right].
     \label{norm_Arr0b_out_exp}
    \end{eqnarray}
     \label{norm_A0_an_exp_out}
\end{subequations}
where $f(R,Z)$ is defined as 
\begin{equation}
    f(R,Z)=-\frac{2(R^2-\alpha^2)}{3(1-Z^2)}.
    \label{f_out}
\end{equation}
We remind the reader that $\mathcal{A}_{\mathit{zz}}^{\mathit{ref}}(Z)$, $\mathcal{A}_{\mathit{\theta \theta}}^{\mathit{ref}}(Z)$, $\mathcal{A}_{\mathit{rz}}^{\mathit{ref}}(Z)$, and $\mathcal{A}_{\mathit{rr}}^{\mathit{ref}}(Z)$  are defined at $R=\alpha$  and, in general, may depend on the Deborah number.

Using (\ref{ND_gov_low_bt_out}) and recalling that $U_{r,0}=\partial U_{r,0}/\partial R=0$ at $Z=\pm1$, it follows that the conformation tensor components at the plates take the form
\begin{equation}
\mathcal{A}_{\theta\theta,0}^{\it{plate}}=0,\quad \mathcal{A}_{\mathit{zz},0}^{\it plate}=1,\quad \mathcal{A}_{\mathit{rz},0}^{\it plate}=\mp\frac{3 De}{2R},\quad\mathcal{A}_{\mathit{rr},0}^{\it plate}=\frac{9 De^2}{2R^2}\quad\mathrm{for\:all} \: De.\label{A_plates}
\end{equation}

\subsubsection{Conformation tensor in the low-$De$ limit}\label{sec: CT low De}

Solving the equations for the conformation tensor components (\ref{ND_gov_low_bt_out}) iteratively for  $De\ll 1$, we obtain
\begin{equation}
\mathcal{A}_{\theta\theta,0}=0,\quad \mathcal{A}_{\mathit{zz},0}=1,\quad 
\mathcal{A}_{\mathit{rz},0}=-\frac{3DeZ}{2R},\quad
\mathcal{A}_{\mathit{rr},0}=\frac{9De^2Z^2}{2R^2}.\label{A0_lowDe}
\end{equation}
We note that the low-$De$ expressions (\ref{A0_lowDe}) are consistent with the results obtained in $\mathsection$~\ref{Low-Deborah-number lubrication analysis} from the low-Deborah-number lubrication analysis, which are valid for $\tilde{\beta} \nll 1$. These expressions for the conformation tensor components also satisfy the boundary conditions at the plates, (\ref{A_plates}). However, the components of the conformation tensor at low Deborah numbers, (\ref{A0_lowDe}), may not satisfy the inlet boundary conditions~(\ref{bc_b_out}). 

\subsection{Pressure drop and elastic stress contributions to the pressure drop at the first order in \texorpdfstring{$\tilde{\beta}$}{}}

The ultra-dilute limit, corresponding to $\tilde{\beta}\ll1$, allows us to calculate the pressure drop at $O(\tilde{\beta})$
using the leading-order velocity field and elastic stresses. Substituting (\ref{Asympt_beta_t}) into (\ref{ND pressure_drop_gen}) yields the pressure drop $\Delta P_1$ at the first order in $\tilde{\beta}$, 
\begin{eqnarray}
      \Delta P_1&=&-\frac{3}{2}\ln{\left(\frac{1}{\alpha}\right)}
-\frac{3 }{2De} \int_{\alpha}^{1}\left[ \int_{-1}^{1}Z\mathcal{A}_{\mathit{ rz},0}\mathrm{d}Z \right]\mathrm{d}R
\nonumber \\
&&-\frac{3 }{4De} \int_{-1}^{1}\left[(1-Z^{2})\left( \left.\mathcal{A}_{ \mathit{rr},0}\right|_{R=\alpha}^{R=1}+\int_{\alpha}^{1}\frac{\mathcal{A}_{ \mathit{rr},0}-\mathcal{A}_{\theta\theta,0}}{R}  \mathrm{d}R\right)\right]\mathrm{d}Z,\label{dP1 ultra-dilute} 
\end{eqnarray}
where $\mathcal{A}_{\theta\theta,0}$, $\mathcal{A}_{\mathit{rz},0}$, and $\mathcal{A}_{\mathit{rr},0}$ are given in (\ref{norm_Athth0b_out_exp}), (\ref{norm_Arz0b_out_exp}), and (\ref{norm_Arr0b_out_exp}).

Similar to (\ref{ND pressure_drop_gen}), the pressure drop at $O(\tilde{\beta})$ in (\ref{dP1 ultra-dilute}) comprises contributions from the Newtonian solvent, elastic shear stresses, and elastic normal stresses. The elastic shear stress (ESS) contribution to the pressure drop at $O(\tilde{\beta})$ is
\begin{equation}
     \Delta P_1^{\it ESS}= -\frac{3}{2De}\int_\alpha^1\left[\int_{-1}^1{Z\mathcal{A}_{rz,0}\mathrm{d}Z}\right]\mathrm{d}R, \label{ESS FO}
\end{equation}
whereas the elastic normal stress (ENS) contribution to the pressure drop at $O(\tilde{\beta})$ is
\begin{equation}
     \Delta P_1^{\it ENS}= -\frac{3 }{4De} \int_{-1}^{1}\left[(1-Z^{2})\left( \left.\mathcal{A}_{ \mathit{rr},0}\right|_{R=\alpha}^{R=1}+\int_{\alpha}^{1}\frac{\mathcal{A}_{ \mathit{rr},0}-\mathcal{A}_{\theta\theta,0}}{R}  \mathrm{d}R\right)\right]\mathrm{d}Z. \label{ENS FO}
\end{equation}
Thus, for a given flow rate $q$, the non-dimensional pressure drop between $R=\alpha$ and $R=1$, $\Delta P =\Delta p/(\eta_0 q/2\pi h^3)$, as a function of $De$, $\alpha$, and $\tilde{\beta}\ll 1$ up to $O(\tilde{\beta})$, is given by 
\begin{equation}
    \Delta P=\Delta P_0(\alpha)+\tilde{\beta}\Delta P_1(\alpha,De)+O(\tilde{\beta}^2,\epsilon^2), \label{dP dP0 plus beta_tdP1}
\end{equation}
where the expressions for $\Delta P_0$ and $\Delta P_1$ are given in (\ref{U0 and dP0 ND}$b$) and (\ref{dP1 ultra-dilute}), respectively.

\subsubsection{Pressure drop at $O(\tilde{\beta})$ in the low-$De$ limit}

At low Deborah numbers, we can calculate the pressure drop and the corresponding expressions for elastic normal and shear stress
contributions to the pressure drop at $O(\tilde{\beta})$ using (\ref{A0_lowDe}) and (\ref{dP1 ultra-dilute})--(\ref{ENS FO}).
Substituting (\ref{A0_lowDe}) into (\ref{ESS FO}) and (\ref{ENS FO}) provides the elastic shear and normal stress contributions to the pressure drop at $O(\tilde{\beta})$ in the low-$De$ limit 
\refstepcounter{equation}
$$
\Delta P_1^{\it ESS}=\frac{3}{2}\ln\left(\frac{1}{\alpha}\right)\quad\hbox{and}\quad \Delta P_1^{\it ENS}=\frac{9}{20}De\left(\frac{1}{\alpha^2}-1\right) \quad \text{for} \quad De\ll1.\eqno{(\theequation{a,b})}\label{ESS and ENS lowDe}
$$
Thus, using (\ref{dP1 ultra-dilute}), (\ref{dP dP0 plus beta_tdP1}), and (\ref{ESS and ENS lowDe}), the total pressure drop between $R=\alpha$ and $R=1$ in the low-$De$ limit is 
\begin{eqnarray}
  \Delta P&&= \underset{\text{Solvent stress}}{\underbrace{\frac{3}{2}(1-\tilde{\beta)}\ln\left(\frac{1}{\alpha}\right)}}
  +\underset{\text{Elastic shear stress}}{\underbrace{\frac{3}{2}\tilde{\beta}\ln\left(\frac{1}{\alpha}\right)}}
  +\underset{\text{Elastic normal stress}}{\underbrace{\frac{9}{20}\tilde{\beta}De\left(\frac{1}{\alpha^2}-1\right)}}
  \nonumber \\&  &
 =\frac{3}{2}\ln{\left(\frac{1}{\alpha}\right)}+ \frac{9}{20}\tilde{\beta}De\left(\frac{1}{\alpha^2}-1\right) \quad \text{for} \quad De\ll1,
    \label{PD_ultradilute_lowDe}
\end{eqnarray}
which is consistent with our previous low-$De$ analysis (\ref{total_PD-low De_nonsym}).

\section{Theoretical and finite-element simulation results}\label{Results}

In this section, we present our theoretical and asymptotic results for the pressure drop and elastic stresses of the Oldroyd-B fluid in a radial flow as developed in the previous sections. We also perform two-dimensional axisymmetric numerical simulations using the finite-element software COMSOL Multiphysics to validate the predictions of the theoretical model. The details of the numerical procedure are provided in Appendix~\ref{App B}.

\begin{figure}
 \centerline{\includegraphics[scale=1]{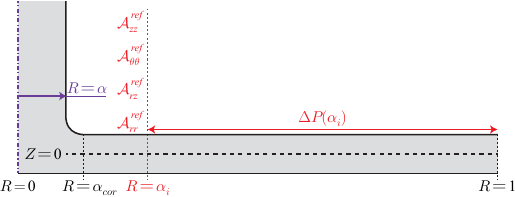}}
\caption{Schematic illustration of the radial-flow configuration used in our analysis. The dimensionless parameter $\alpha_i$ corresponds to the radial location at which we extract the values
of ($\mathcal{A}_{\mathit{zz}}^{\mathit{ref}},\mathcal{A}_{\mathit{\theta \theta}}^{\mathit{ref}}, \mathcal{A}_{\mathit{rz}}^{\mathit{ref}},\mathcal{A}_{\mathit{rr}}^{\mathit{ref}})$ from finite-element simulations. To facilitate the simulations, we round the corner between the inlet tube and the top plate to a quarter-circle of radius $r_{\it cor}=h$, and define the dimensionless parameter $\alpha_{\it cor}=(r_{\it in}+r_{\it cor})/r_{\it out}$. Our primary interest is to
determine the non-dimensional pressure drop $\Delta P(\alpha_i)$ and the spatial variation of elastic stresses.}\label{F2}
\end{figure}

We note that the distributions of $\mathcal{A}_{\mathit{zz}}^{\mathit{ref}}(Z)$, $\mathcal{A}_{\mathit{\theta \theta}}^{\mathit{ref}}(Z)$, $\mathcal{A}_{\mathit{rz}}^{\mathit{ref}}(Z)$, and $\mathcal{A}_{\mathit{rr}}^{\mathit{ref}}(Z)$, introduced in $\mathsection$~\ref{PF}, are unknown \emph{a priori}, and are difficult to determine theoretically. Therefore, we extract these distributions from finite-element simulations.
To this end, we introduce the dimensionless parameter $\alpha_i$ to denote the radial location at which we extract the values of $\mathcal{A}_{\mathit{zz}}^{\mathit{ref}}(Z)$, $\mathcal{A}_{\mathit{\theta \theta}}^{\mathit{ref}}(Z)$, $\mathcal{A}_{\mathit{rz}}^{\mathit{ref}}(Z)$, and $\mathcal{A}_{\mathit{rr}}^{\mathit{ref}}(Z)$. We then present the results for the pressure drop $\Delta P(\alpha_i)$, defined as $\Delta P(\alpha_i)=P(R=\alpha_i)-P(R=1)$, as shown in figure~\ref{F2}. Knowing the values of ($\mathcal{A}_{\mathit{zz}}^{\mathit{ref}},\mathcal{A}_{\mathit{\theta \theta}}^{\mathit{ref}}, \mathcal{A}_{\mathit{rz}}^{\mathit{ref}},\mathcal{A}_{\mathit{rr}}^{\mathit{ref}})$ at $R=\alpha_i$, we determine the conformation tensor components in the region $\alpha_i \leq R \leq 1$ using (\ref{norm_A0_an_exp_out}). For given values of $\alpha_i$, $De$, and $\tilde{\beta}\ll1$, we then use MATLAB’s routine \texttt{trapz} to calculate the dimensionless pressure drop $\Delta P(\alpha_i)$ in the ultra-dilute limit via (\ref{dP dP0 plus beta_tdP1}), with $\alpha$ replaced by $\alpha_i$. Typical values of the grid size are $\Delta R  = 10^{-4}$ and $\Delta Z =10^{-2}$.

\subsection{Radial variation of the conformation tensor components between the plates and their dependence on the Deborah number}

\begin{figure}
\centerline{\includegraphics[scale=1]{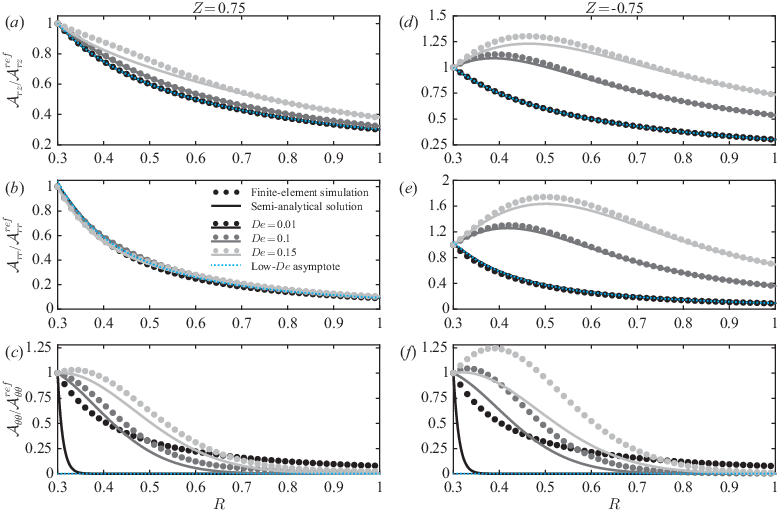}}
\caption{The radial variation of conformation tensor components on ($a$--$c$) $Z=0.75$ and ($d$--$f$) $Z=-0.75$ in a radial flow of the Oldroyd-B fluid between parallel plates in the ultra-dilute limit. Scaled elastic stresses $(a,d)$ $\mathcal{A}_{\mathit{rz}}/\mathcal{A}_{\it rz}^{\it{ref}}$, $(b,e)$  $\mathcal{A}_{ \mathit{rr}}/\mathcal{A}_{\it rr}^{\it{ref}}$, and $(c,f)$ $\mathcal{A}_{\theta \theta}/\mathcal{A}_{\it \theta\theta}^{\it{ref}}$ as a function of $R$ for $De=0.01,0.1$, and $0.15$. Dots represent the results of the finite-element simulation.
Solid lines represent the leading-order semi-analytical solutions (\ref{norm_Athth0b_out_exp}), (\ref{norm_Arz0b_out_exp}), and (\ref{norm_Arr0b_out_exp}).  Cyan dotted lines represent the low-$De$ asymptotic solution (\ref{A0_lowDe}). All calculations were performed using $\tilde{\beta}=0.05$ and $\alpha_{i}=0.3$.}
\label{F3}
\end{figure}

First, we examine the spatial variation of the elastic stresses and their dependence on the Deborah number. Figure \ref{F3} presents the radial variation of the elastic stresses $\mathcal{A}_{\mathit{rz}}$,  $\mathcal{A}_{ \mathit{rr}}$, and  $\mathcal{A}_{\theta \theta}$, scaled by the reference values extracted from finite-element simulations at $\alpha_i=0.3$, for ($a$--$c$) $Z=0.75$ and ($d$--$f$) $Z=-0.75$, in a radial flow of the Oldroyd-B fluid between parallel plates in the ultra-dilute limit. Solid lines represent the semi-analytical solutions (\ref{norm_Athth0b_out_exp}), (\ref{norm_Arz0b_out_exp}), and (\ref{norm_Arr0b_out_exp}). Dots represent the results of the finite-element simulation. Cyan dotted lines represent the low-$De$ asymptotic solution (\ref{A0_lowDe}).

For a small Deborah number of $De=0.01$, the elastic stresses are symmetric about the midplane $Z=0$ and monotonically decrease with increasing $R$. Specifically, the elastic shear stress relaxes as $R^{-1}$ (figure~\ref{F3}$(a,d)$), whereas the elastic radial normal stress relaxes as $R^{-2}$ (figure~\ref{F3}$(b,e)$), consistent with the low-$De$ asymptotic solution (\ref{A0_lowDe}). Furthermore, there is excellent agreement
between the semi-analytical and finite-element simulation results.
It is evident from figure~\ref{F3}$(c,f)$ that, for $De=0.01$, both the ultra-dilute theory and the finite-element simulations predict a rapid decrease in the scaled elastic hoop stress, $\mathcal{A}_{\theta \theta}/\mathcal{A}_{\it \theta\theta}^{\it{ref}}$, from its inlet value of unity at $\alpha_i  =0.3$ to its fully relaxed value of zero downstream. However, the simulations show a noticeably slower spatial relaxation than the theoretical prediction.
We note that the low-$De$ asymptotic solution (\ref{A0_lowDe}) does not capture the rapid initial decay of $\mathcal{A}_{\theta \theta}/\mathcal{A}_{\it \theta\theta}^{\it{ref}}$ from its value of unity at $\alpha_i = 0.3$, owing to its failure to satisfy the inlet boundary condition (\ref{bc_b_out}).

For intermediate Deborah numbers ($De=0.1$ and $De=0.15$), the elastic stresses become increasingly asymmetric about the midplane. We observe that, at $Z=0.75$, the elastic stresses exhibit a monotonic relaxation. In contrast, at  $Z=-0.75$, the elastic stresses exhibit a non-monotonic variation, first increasing with $R$, reaching a maximum, and then decreasing with $R$. Furthermore, as expected, both the growth and relaxation of the elastic stresses occur over a longer length scale as the Deborah number increases. Nevertheless, for $De=0.1$ and $De=0.15$, the ultra-dilute theory captures both the asymmetry and the non-monotonic behavior of the elastic stresses fairly well, showing good agreement with the finite-element simulation results.

\begin{figure}
\centerline{\includegraphics[scale=1]{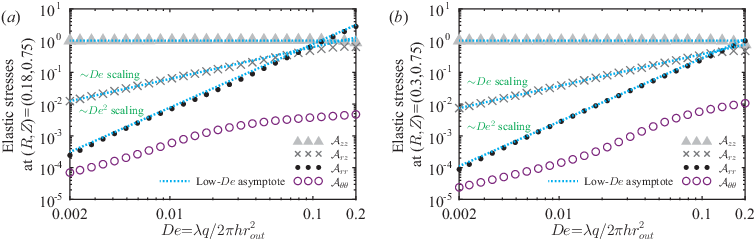}}
\caption{$(a,b)$ Values of the conformation tensor components $\mathcal{A}_{\mathit{zz}}$, $\mathcal{A}_{\mathit{rz}}$, $\mathcal{A}_{\mathit{rr}}$, and $\mathcal{A}_{\mathit{\theta\theta}}$ as a function of $De=\lambda q/(2 \pi h r_{\it out}^2)$, evaluated at $(a)$ $(R,Z)=(0.18,0.75)$ and $(b)$ $(R,Z)=(0.3,0.75)$. Triangles, crosses, dots, and circles represent the results of the finite-element simulation. Cyan dotted lines represent the low-$De$ asymptotic solution (\ref{A0_lowDe}). All calculations were performed using $\tilde{\beta}=0.05$.}
\label{F4}
\end{figure}

The low-$De$ analysis (\ref{A0_lowDe}) and the predicted elastic stresses on the plates (\ref{A_plates}) indicate that $\mathcal{A}_{\it zz}$ and $\mathcal{A}_{\it \theta\theta}$ are independent of $De$, while $\mathcal{A}_{\it rz}$ and $\mathcal{A}_{\it rr}$ scale as $De$ and $De^2$, respectively. To test these predictions and further elucidate the behavior of the elastic stresses, we present in figure~\ref{F4}($a,b$) the conformation tensor components as a function of $De$, evaluated at $(a)$ $(R,Z)=(0.18,0.75)$ and $(b)$ $(R,Z)=(0.3,0.75)$, for $\tilde{\beta}=0.05$. Triangles, crosses, dots, and circles represent the results of the finite-element simulation. Cyan dotted lines represent the low-$De$ asymptotic solution (\ref{A0_lowDe}). 

Consistent with the low-$De$ asymptotic solution, $\mathcal{A}_{\it zz}$ exhibits only a weak dependence on the Deborah number, while $\mathcal{A}_{\it rz}$ and $\mathcal{A}_{\it rr}$ scale linearly and quadratically with $De$, respectively, for $De \lesssim 0.1$. For higher Deborah numbers, $\mathcal{A}_{\it rz}$ no longer follows the predicted linear scaling, whereas $\mathcal{A}_{\it rr}$ retains its quadratic dependence on $De$ throughout the investigated range of
Deborah numbers. 
While the low-$De$ analysis predicts that the elastic hoop stress, represented by $\mathcal{A}_{\theta\theta}$ and associated with polymer stretching in the circumferential direction, is identically zero, figure~\ref{F4}$(a,b)$ clearly shows that $\mathcal{A}_{\theta\theta}$ is small but non-zero, consistent with the results shown in figures~\ref{F3}$(c,f)$. Moreover, $\mathcal{A}_{\theta\theta}$ monotonically increases with $De$.

\subsection{Variation of pressure drop with the Deborah number in the ultra-dilute limit}\label{dP ultra-dilute}

\begin{figure}
\centerline{\includegraphics[scale=1]{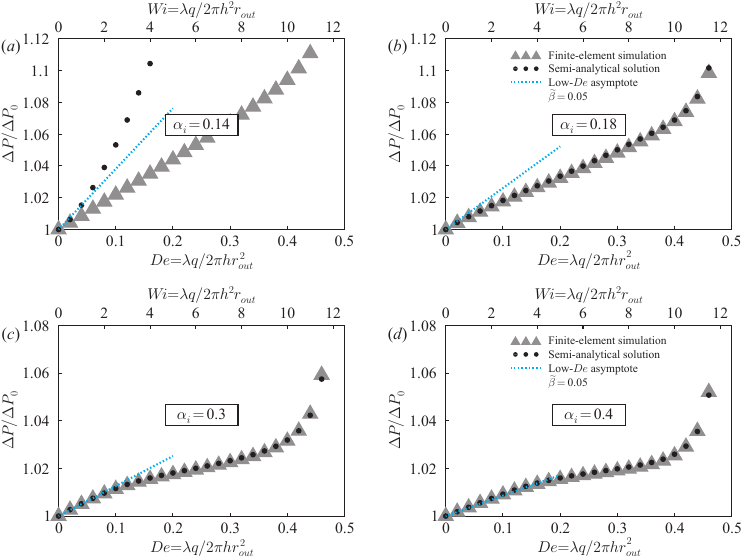}}
\caption{Non-dimensional pressure drop of the Oldroyd-B fluid in a
radial flow between parallel plates in the ultra-dilute limit. ($a$--$d$) Scaled pressure drop $\Delta P/\Delta P_0$ as a function of $De=\lambda q/(2 \pi h r_{\it out}^2)$ (or $Wi=\lambda q/(2 \pi h^2 r_{\it out})$) for $(a)$ $\alpha_i=0.14$, $(b)$ $\alpha_i=0.18$, ($c$) $\alpha_i=0.3$, and ($d$) $\alpha_i=0.4$. Gray triangles represent the results of the finite-element
simulation. Black dots represent the semi-analytical solution (\ref{dP dP0 plus beta_tdP1}). Cyan dotted lines represent the low-$De$ asymptotic solution (\ref{PD_ultradilute_lowDe}).
All calculations were performed using $\tilde\beta=0.05$.}\label{F5}
\end{figure}

In this subsection, we study the pressure drop of the Oldroyd-B fluid in a radial flow between parallel plates at order-one Deborah numbers in the ultra-dilute limit, $\tilde{\beta}=0.05$.
We present in figure~\ref{F5}($a$--$d$) the scaled pressure drop $\Delta P/\Delta P_0$ as a function of $De=\lambda q/(2 \pi h r_{\it out}^2)$ (or $Wi=\lambda q/(2 \pi h^2 r_{\it out})$) for the radial flow of an Oldroyd-B fluid in the ultra-dilute limit for $(a)$ $\alpha_i=\alpha_{\it cor}=0.14$, $(b)$ $\alpha_i=0.18$, $(c)$ $\alpha_i=0.3$, and $(d)$ $\alpha_i=0.4$. Gray triangles represent the finite-element simulation results obtained from calculating the pressure drop along the midplane $Z =0$. Black dots represent the semi-analytical solution (\ref{dP dP0 plus beta_tdP1}) based on the low-$\tilde{\beta}$ lubrication analysis. Cyan dotted lines represent the low-$De$ asymptotic solution (\ref{PD_ultradilute_lowDe}).

It is evident from figure~\ref{F5}$(a)$ that for $\alpha_i=\alpha_{\it cor}=0.14$, neither the semi-analytical predictions nor the low-$De$ asymptotic solution accurately captures the pressure drop, even at low Deborah numbers, due to significant entrance effects. However, when $\alpha_i$ increases to $0.18$, we observe excellent agreement
between the semi-analytical and finite-element simulation results throughout the investigated range of
Deborah numbers, as shown in figure~\ref{F5}$(b)$. 
Nevertheless, for $\alpha_i=0.18$, the low-$De$ asymptotic solution still fails to accurately predict the pressure drop. This discrepancy arises because, in this case, the low-$De$ conformation tensor components (\ref{A0_lowDe}) do not satisfy the inlet boundary conditions~(\ref{bc_b_out}). 

When $\alpha_i$ is further increased to $0.3$ and $0.4$, there is excellent agreement between the semi-analytical and simulation results, as shown in figure~\ref{F5}($c,d$).
Furthermore, for $\alpha_i=0.3$ and $\alpha_i=0.4$, we find that the low-$De$ asymptotic solution accurately captures both the semi-analytical predictions and the simulation results at low Deborah numbers, and that the range of good agreement increases with $\alpha_i$.

\begin{figure}
\centerline{\includegraphics[scale=1]{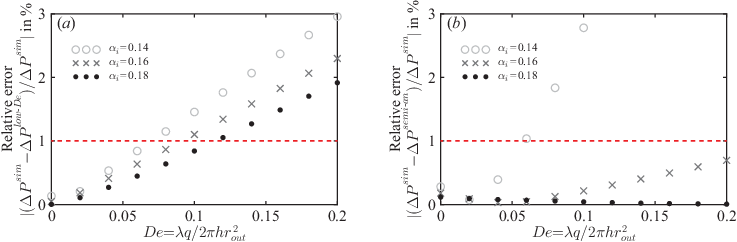}}
\caption{($a$) Relative error between the low-$De$ asymptotic solution (\ref{PD_ultradilute_lowDe}) and numerical simulations for the pressure drop $\Delta P(\alpha_i)$, $|(\Delta P^{ \it sim}-\Delta P^{ \it low\text{-}De})/\Delta P^{ \it sim}|$, as a function of $De$ for $\alpha_i=0.14$ (circles), $0.16$ (crosses), and $0.18$ (dots). ($b$) Relative error between the semi-analytical solution (\ref{dP dP0 plus beta_tdP1}) and numerical simulations for the pressure drop $\Delta P(\alpha_i)$, $|(\Delta P^{ \it sim}-\Delta P^{ \it semi\text{-}an})/\Delta P^{ \it sim}|$, as a function of $De$ for $\alpha_i=0.14$ (circles), $0.16$ (crosses), and $0.18$ (dots). Red dashed lines represent the relative error of 1 \%. All calculations were performed using $\tilde\beta=0.05$.}\label{F6}
\end{figure}

To further elucidate the weak agreement between the low-$De$ predictions and the simulation results at relatively small values of $\alpha_i$, as shown in figure~\ref{F5}$(a,b)$, we present in figure~\ref{F6}($a$) the relative error in the pressure drop between the low-$De$ asymptotic solution and numerical simulations as a function of $De$ for $\alpha_i=0.14,0.16$, and $0.18$.
Figure~\ref{F6}($b$) shows the corresponding relative error between the semi-analytical solution and numerical simulations for the pressure drop as a function of $De$ for $\alpha_i=0.14,0.16$, and $0.18$. We observe that for $\alpha_i=0.14$, the relative errors, defined as $|(\Delta P^{ \it sim}-\Delta P^{ \it low\text{-}De})/\Delta P^{ \it sim}|$ and $|(\Delta P^{ \it sim}-\Delta P^{ \it semi\text{-}an})/\Delta P^{ \it sim}|$, respectively, exceed 1 \% even for $De<0.1$. However, the relative error of the semi-analytical solution decreases rapidly with $\alpha_i$, and for $\alpha_i=0.16$, it remains below 0.7 \% up to $De=0.2$. In contrast, as shown in figure~\ref{F6}$(a)$, the relative error of the low-$De$ asymptotic solution decreases more slowly. For example, for $\alpha_i=0.18$, the relative error is approximately 2 \% for up to $De=0.2$. Therefore, the low-$De$ asymptotic solution agrees well with the simulation results only at larger values of $\alpha_i$, as shown in figure~\ref{F5}($c,d$).

Consistent with previous studies on radial flows of weakly viscoelastic fluids~\citep{lee_radial_1976,co1977slow}, our semi-analytical and numerical simulation results in figure~\ref{F5} predict a monotonic increase in the
pressure drop of the Oldroyd-B fluid with increasing $De$ (or $Wi$). In the next subsection, we provide insight into the mechanisms underlying this increase in pressure drop.

\subsection{Different contributions to the pressure drop between the plates}

\begin{figure}
\centerline{\includegraphics[scale=1]{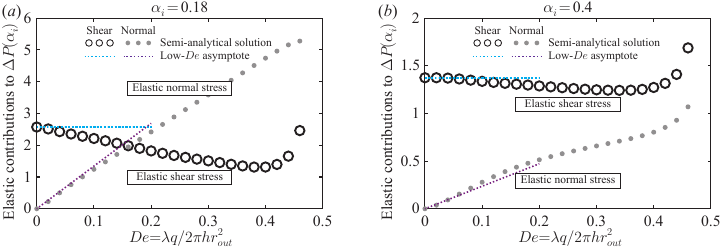}}
\caption{Elastic contributions to the non-dimensional pressure drop of the Oldroyd-B fluid, scaled by $\tilde{\beta}$, as a function of $De=\lambda q/(2 \pi h r_{\it out}^2)$ in a radial flow between parallel plates for $(a)$ $\alpha_i=0.18$ and $(b)$ $\alpha_i=0.4$ in the ultra-dilute limit. Black circles and gray dots represent the semi-analytical solutions (\ref{ESS FO}) and (\ref{ENS FO}) for elastic shear and normal stress contributions.
Cyan dotted and purple lines represent the low-$De$ asymptotic solution (\ref{ESS and ENS lowDe}) for elastic shear and normal stress contributions.}\label{F7}
\end{figure}

It follows from figure~\ref{F5}($a$--$d$) that the pressure drop of the Oldroyd-B fluid in a radial flow monotonically increases with $De$. To elucidate the origin of this increase in pressure drop, we examine the relative importance of elastic contributions to the pressure drop. We present in figure~\ref{F7}$(a,b)$ the elastic contributions to the non-dimensional pressure drop $\Delta P (\alpha_i)$, scaled by $\tilde{\beta}$, as a function of $De=\lambda q/(2 \pi h r_{\it out}^2)$ for $(a)$ $\alpha_i=0.18$ and $(b)$ $\alpha_i=0.4$ in the ultra-dilute limit.  Black circles and gray dots represent the elastic shear and normal stress contributions obtained from the semi-analytical solutions (\ref{ESS FO}) and (\ref{ENS FO}). Cyan and purple dotted lines represent the elastic shear and normal stress contributions obtained from the low-$De$ asymptotic solutions (\ref{ESS and ENS lowDe}$a$) and (\ref{ESS and ENS lowDe}$b$).

For both $\alpha_i=0.18$ and $\alpha_i=0.4$, we observe that the elastic normal stress contribution to the pressure drop monotonically increases with $De$. In contrast, the elastic shear stress contribution first decreases, reaches a minimum at $De \approx 0.4$, and then increases with $De$. However, the variation in the elastic shear stress contribution is weaker than that in the elastic normal stress contribution. Thus, we conclude that the primary mechanism for the increased pressure drop between the plates is the high elastic normal stresses upstream, in contrast to those much farther downstream. The flow requires a higher pressure drop to overcome these large upstream elastic normal stresses and drive the fluid between the plates.

It is evident from figure~\ref{F7}$(a,b)$ that the low-$De$ asymptotic solution (\ref{ESS and ENS lowDe}$b$) for the elastic normal stress contribution accurately captures the semi-analytical prediction for both $\alpha_i=0.18$ and $\alpha_i=0.4$ at low Deborah numbers.
However, the corresponding low-$De$ asymptotic solution (\ref{ESS and ENS lowDe}$a$) for the elastic shear stress contribution captures fairly well the semi-analytical prediction only for $\alpha_i=0.4$. This discrepancy is consistent with the weak agreement between the low-$De$ asymptotic solution~(\ref{PD_ultradilute_lowDe}) and the semi-analytical prediction for the pressure drop shown in figure~\ref{F5}$(b)$ for $\alpha_i=0.18$, indicating that the main source of this disagreement is the elastic shear stress contribution. 

\subsection{Assessing the effect of the viscosity ratio on the pressure drop and radial velocity}

\begin{figure}
\centerline{\includegraphics[scale=1]{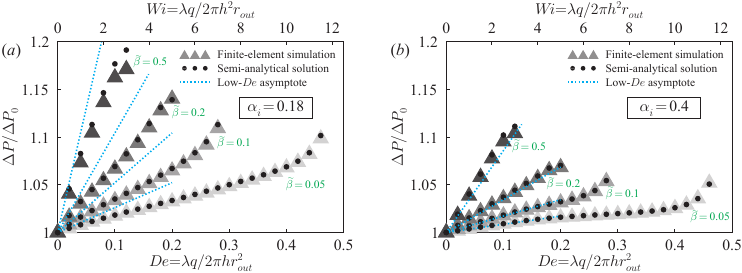}}
\caption{The effect of the polymer-to-total viscosity ratio on the pressure drop of the Oldroyd-B fluid in a radial flow between parallel plates. ($a,b$) Scaled pressure drop $\Delta P/\Delta P_0$ as a function of $De=\lambda q/(2 \pi h r_{\it out}^2)$ (or $Wi=\lambda q/(2 \pi h^2 r_{\it out})$) for $(a)$ $\alpha_i=0.18$ and $(b)$ $\alpha_i=0.4$ and different values of the viscosity ratio $\tilde{\beta}=0.05$, $0.1$, $0.2$, and $0.5$. Triangles represent the results of the finite-element
simulation. Black dots represent the semi-analytical solution (\ref{dP dP0 plus beta_tdP1}). Cyan dotted lines represent the low-$De$ asymptotic solution (\ref{total_PD-low De_nonsym}).}
\label{F8}
\end{figure}

In previous subsections, we presented the results for the dimensionless pressure drop of the Oldroyd-B fluid in a radial flow in the ultra-dilute limit, with $\tilde{\beta}=0.05$. Specifically, we showed excellent agreement between the semi-analytical ultra-dilute predictions and the finite-element simulation results. To assess the range of validity of our ultra-dilute theory and elucidate the effect of the polymer-to-total viscosity ratio on the pressure drop between the plates, we present in figure~\ref{F8}$(a,b)$ the scaled pressure drop $\Delta P/\Delta P_0$ as a function of $De=\lambda q/2\pi h r_{\it out}^2$ (or $Wi= \lambda q/2\pi h^2 r_{\it out}$) for $(a)$ $\alpha_i=0.18$ and $(b)$ $\alpha_i=0.4$ and four different values of the viscosity ratio $\tilde{\beta}=0.05$, $0.1$, $0.2$, and $0.5$. Triangles represent the results of the finite-element simulation, black dots represent the semi-analytical solution, and cyan dotted lines represent the low-$De$ asymptotic solution (\ref{total_PD-low De_nonsym}).

First, we observe that the pressure drop of the Oldroyd-B fluid in a radial flow monotonically increases with $De$ for all values of $\tilde{\beta}$, consistent with results shown in figure~\ref{F5}. Second, similar to the case of $\tilde{\beta}=0.05$, there is excellent agreement between the predictions of our ultra-dilute theory and the finite-element simulation results when $\tilde{\beta}=0.1$ and $\tilde{\beta}=0.2$, for both $\alpha_i=0.18$ and $\alpha_i=0.4$, across the entire range of Deborah numbers considered. As expected, with increasing $\tilde{\beta}$, the agreement between the ultra-dilute theory and the simulations deteriorates. Nevertheless, even for $\tilde{\beta}=0.5$, which clearly does not lie within the ultra-dilute limit $\tilde{\beta}\ll1$, the relative error remains below $2$ \% up to $De=0.12$. In COMSOL Multiphysics, we are currently unable to obtain converged finite-element simulation results beyond $De=0.12$ for $\tilde{\beta}=0.5$. Therefore, further investigation would be required to assess the pressure drop of an Oldroyd-B fluid in a radial flow between parallel plates at higher Deborah numbers.

Furthermore, similar to the results shown in figure~\ref{F5}($b$), we observe in figure~\ref{F8}$(a)$ that, for $\alpha_i=0.18$, the low-$De$ asymptotic solution (\ref{total_PD-low De_nonsym}) (or (\ref{PD_ultradilute_lowDe})) fails to accurately predict the pressure drop, even at
low Deborah numbers. As noted in \S~\ref{dP ultra-dilute}, this failure arises because the low-$De$ asymptotic solution relies on the low-$De$ conformation tensor components (\ref{lowDe_A_1_nonsym}) and (\ref{Arr2_nonsym}) (or (\ref{A0_lowDe})), which do not satisfy the inlet boundary conditions (\ref{bc_b_out}). However, for $\alpha_i=0.4$, the low-$De$ conformation tensor components comply with the inlet boundary conditions (\ref{bc_b_out}), and therefore the low-$De$ asymptotic solution accurately captures the pressure drop behavior for all values of $\tilde{\beta}$, consistent with the results shown in figure~\ref{F5}($d$).

\begin{figure}
\centerline{\includegraphics[scale=1]{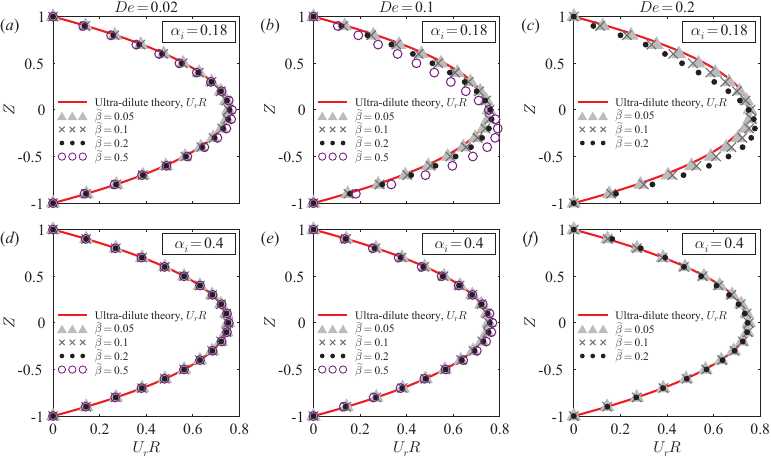}}
\caption{Non-dimensional radial velocity distribution of the Oldroyd-B fluid in a radial flow between parallel plates for different values of the polymer-to-total viscosity ratio. ($a$--$f$) Radial velocity $U_r$ multiplied by the radial coordinate $R$, $U_r R$, as a function of $Z$ for $De=0.02, 0.1$, and $0.2$ and different values of the viscosity ratio $\tilde{\beta}=0.05$, $0.1$, $0.2$, and $0.5$, with ($a$--$c$) $R=\alpha_i=0.18$ and ($d$--$f$) $R=\alpha_i=0.4$. Red solid lines represent the ultra-dilute theoretical prediction $U_r R=(3/4)(1-Z^2)$. Triangles, crosses, dots, and circles represent the results of the finite-element simulation.}
\label{F9}
\end{figure}

In addition to the pressure drop, it is of particular interest to elucidate the effect of the polymer-to-total viscosity ratio on the radial velocity. To this end, we present in figure~\ref{F9}($a$--$f$) the scaled radial velocity $U_r R$ as a function of $Z$ for $De=0.02, 0.1$, and $0.2$ and different values of the viscosity ratio $\tilde{\beta}=0.05$, $0.1$, $0.2$, and $0.5$, with ($a$--$c$) $R=\alpha_i=0.18$ and ($d$--$f$) $R=\alpha_i=0.4$. Red solid lines represent our ultra-dilute theoretical prediction $U_r R=(3/4)(1-Z^2)$ and triangles, crosses, dots, and circles represent the finite-element simulation results.

As expected, for a small Deborah number of $De=0.02$, the radial velocity at $R=\alpha_i=0.18$ is symmetric and parabolic, in excellent agreement with the theoretical prediction for all values of $\tilde{\beta}$ (figure~\ref{F9}($a$)). However, as $De$ increases to $0.1$ and $0.2$, the radial velocity loses its symmetry about the midplane $Z=0$ for all values of $\tilde{\beta}$ except $\tilde{\beta}=0.05$, and slightly deviates from the theoretical profile (figure~\ref{F9}($b,c$)). Clearly, both the asymmetry and the deviation from the theoretical profile become more pronounced with increasing $\tilde{\beta}$ and $De$. Nevertheless, the excellent agreement between the ultra-dilute theory, based on the parabolic velocity profile $U_r R=(3/4)(1-Z^2)$, and the finite-element simulations shown in figure~\ref{F8}$(a)$ suggests that the observed asymmetry and deviation from the parabolic profile have only a minor effect on the resulting pressure drop. 
Furthermore, we observe that as $R$ increases to $R=\alpha_i=0.4$, the radial velocity remains approximately symmetric and parabolic, in excellent agreement with the theoretical prediction for all considered values of $De$, namely $De=0.02,0.1$, and $0.2$, and for all values of $\tilde{\beta}$ (figure~\ref{F9}($d$--$f$)).

\section{Concluding remarks}\label{CR}

In this work, we studied the pressure-driven radial flow of an Oldroyd-B fluid between two parallel disc-shaped plates. 
We applied the lubrication approximation and presented a
theoretical framework for calculating the dimensionless pressure drop between the plates. Employing the low-Deborah-number
lubrication analysis, we provided an analytical expression for the non-dimensional pressure drop of the Oldroyd-B up to $O(De)$, complementing the earlier analyses of \citet{lee_radial_1976} and \citet{co1977slow}, who used the five-constant Oldroyd and third-order fluid models, respectively. We exploited the one-way coupling between the Newtonian velocity field and the elastic stresses in the ultra-dilute limit, allowing us to derive semi-analytical expressions for the conformation tensor and pressure drop, and to elucidate the pressure drop behavior at order-one Deborah numbers. We performed finite-element numerical simulations to complement the theoretical analysis and validate the predicted pressure drop and spatial distribution of elastic stresses between the plates, finding excellent agreement with the theoretical predictions.

The pressure drop of an Oldroyd-B fluid in a radial-flow configuration monotonically increases with $De$, as shown in figure~\ref{F5}. To elucidate the source of this pressure drop behavior, we examined the relative importance of elastic normal and shear stress contributions to the pressure drop (see figure~\ref{F7}). 
Our analysis revealed that the primary mechanism responsible for the increase in pressure drop is the development of high elastic normal stresses upstream compared to the corresponding values downstream. As a result, a higher pressure drop is required to overcome these elevated upstream elastic stresses and drive the fluid between the plates.
In addition, we delineated the range of validity of the ultra-dilute approximation by performing numerical simulations for four different values of the viscosity ratio $\tilde{\beta}=0.05$, $0.1$, $0.2$, and $0.5$. We found that even for $\tilde{\beta}=0.5$, which does not satisfy the assumption of the ultra-dilute limit $\tilde{\beta}\ll1$, the predictions of the ultra-dilute theory remain in good agreement with the numerical results. 

Our results indicate that the low-$De$ asymptotic solution (\ref{total_PD-low De_nonsym}) for the pressure drop of an Oldroyd-B fluid, which is consistent with previous results of~\citet{lee_radial_1976} and \cite{co1977slow}, does not accurately predict the pressure drop between the plates in all cases (see figure~\ref{F5}($b$) with $\alpha_i=0.18$) because the associated conformation tensor components fail to satisfy the inlet boundary conditions.
In contrast, our theoretical approach, based on lubrication theory and the ultra-dilute limit, incorporates the inlet boundary conditions for the conformation tensor components and accurately predicts the pressure drop of the Oldroyd-B fluid, thus allowing us to study the pressure drop behavior at order-one Deborah numbers. We, therefore, believe that our theoretical results in the ultra-dilute limit at $De=O(1)$ are of fundamental interest, as they can be helpful for validating numerical simulations and advancing our understanding of viscoelastic radial flows.

Our semi-analytical approach relies on knowledge of the conformation tensor components $\mathcal{A}_{\mathit{zz}}^{\mathit{ref}}(Z)$, $\mathcal{A}_{\mathit{\theta \theta}}^{\mathit{ref}}(Z)$, $\mathcal{A}_{\mathit{rz}}^{\mathit{ref}}(Z)$, and $\mathcal{A}_{\mathit{rr}}^{\mathit{ref}}(Z)$ at $R=\alpha_i$ for different Deborah numbers, which are obtained from finite-element simulations. Unfortunately, in COMSOL Multiphysics, we are currently unable to obtain converged finite-element simulation results beyond $De=0.46$ ($Wi=11.5$) for $\tilde{\beta}=0.05$. Therefore, as a future research direction, it would be interesting to obtain numerical results at higher Deborah numbers using the log-conformation formulation~\citep{fattal2004constitutive}, as implemented, for example, within the finite-volume software OpenFOAM~\citep{jasak2007openfoam}, coupled with the viscoelastic flow solver RheoTool~\citep{pimenta2017stabilization}, similar to recent studies by~\citet{mahapatra2025viscoelastic} and~\citet{kedem2026viscoelastic}.

Finally, we note that it would be interesting to compare our theoretical predictions for the flow rate--pressure drop relation in a radial flow of an Oldroyd-B fluid with experimental measurements. However, to the best of our knowledge, experimental data on the flow rate--pressure drop relation for constant shear-viscosity Boger fluids in radial flows remain limited. Indeed, the working fluid employed in the experiments of \citet{lee_radial_1976_exp} exhibited both strong elastic and shear-thinning effects, making it unsuitable for a direct comparison with the present theory. We, therefore, believe that further experimental investigations are needed to enable a quantitative comparison with our theoretical framework.

\backsection[Funding]{We gratefully acknowledge support by the Israel Science Foundation (Grant No. 1942/23). 
E.B.\ acknowledges the support from the Israeli Council for Higher Education Yigal Alon Fellowship.}

\backsection[Declaration of interests]{The authors report no conflict of interest.}

\backsection[Author ORCIDs]{
\\Ron Lottem https://orcid.org/0009-0000-3047-0773;\\
Evgeniy Boyko https://orcid.org/0000-0002-9202-5154.}

\appendix

\section{Details of the non-dimensionalization}\label{App A}

In this appendix, we provide additional details for the non-dimensionalization of our governing equations in the lubrication limit.  Under the assumption that $\boldsymbol{u}=u_{r}\boldsymbol{e}_r$, the governing equations (\ref{Continuity+Momentum}$b$)--(\ref{A}) in the dimensional form are given by 
\begin{equation}
\frac{\partial p}{\partial r}=\eta_s\frac{\partial^{2}u_{r}}{\partial z^{2}}+\frac{\eta_p}{\lambda}\frac{1}{r}\frac{\partial(r(A_{ \it rr}-1))}{\partial r}+\frac{\eta_p}{\lambda}\frac{\partial A_{\it zr}}{\partial z}-\frac{\eta_p}{\lambda}\frac{A_{\theta\theta}-1}{r},\label{Momentum-r Cyl}
\end{equation}
\begin{equation}
\frac{\partial p}{\partial z}=\frac{\eta_p}{\lambda}\frac{\partial A_{\it zz}}{\partial z}+\frac{\eta_p}{\lambda}\frac{1}{r}\frac{\partial(rA_{\it rz})}{\partial r},\label{Momentum-z Cyl}
\end{equation}
\begin{equation}\label{Dim Azz}
u_{r}\frac{\partial A_{\it zz}}{\partial r}=-\frac{1}{\lambda}(A_{\it zz}-1),
\end{equation}
\begin{equation}\label{Dim Athth} 
u_{r}\frac{\partial A_{\theta\theta}}{\partial r}-2\frac{u_{r}}{r}A_{\theta\theta}=-\frac{1}{\lambda}(A_{\theta\theta}-1),
\end{equation}
\begin{equation}\label{Dim Arz} 
u_{r}\frac{\partial A_{\it rz}}{\partial r}-\frac{\partial u_{r}}{\partial z} A_{\it zz}+\frac{u_{r}}{r}A_{\it rz}=-\frac{1}{\lambda}A_{\it rz},
\end{equation}
\begin{equation}\label{Dim Arr}
u_{r}\frac{\partial A_{\it rr}}{\partial r}-2\frac{\partial u_{r}}{\partial z} A_{\it rz}-2\frac{\partial u_{r}}{\partial r} A_{\it rr}=-\frac{1}{\lambda}(A_{\it rr}-1).
\end{equation}
Substituting the non-dimensional variables (\ref{ND_variables})--(\ref{De and Wi}) into the governing equations (\ref{Momentum-r Cyl})--(\ref{Dim Arr}) leads to
\begin{equation}
\frac{\partial P}{\partial R}=(1-\tilde{\beta})\frac{\partial^2 U_{r}}{\partial Z^2}+\frac{\tilde{\beta}}{De}\left(\frac{1}{R}\frac{\partial (R(\mathcal{A}_{\it rr}-\epsilon^2))}{\partial R}+\frac{\partial \mathcal{A}_{\it rz}}{\partial Z}-\frac{\mathcal{A}_{\theta\theta}-\epsilon^2}{R}\right),
\label{dP_dR_ND long}
\end{equation}
\begin{equation}
\frac{\partial P}{\partial Z}=\epsilon^2\frac{\tilde{\beta}}{De}\left(\frac{\partial \mathcal{A}_{\it zz}}{\partial Z}+\frac{1}{R}\frac{\partial (R\mathcal{A}_{\it rz})}{\partial R}\right),
\label{dP_dZ_ND long}      
\end{equation}
\begin{equation}
U_{r} \frac{\partial \mathcal{A}_{\it zz}}{\partial R}= -\frac{1}{De}( \mathcal{A}_{\it zz}-1),
\label{Azz_ND long}
\end{equation}
\begin{equation}
U_{r}\frac{\partial \mathcal{A}_{\theta\theta}}{\partial r}-2\frac{U_{r}}{R}\mathcal{A}_{\theta\theta}=-\frac{1}{De}(\mathcal{A}_{\theta\theta}-\epsilon^2), \label{Athth ND long} 
\end{equation}
\begin{equation}
U_{r} \frac{\partial \mathcal{A}_{\it rz}}{\partial R}-\frac{\partial U_{r}}{\partial Z} \mathcal{A}_{\it zz}+\frac{U_{r}}{R}\mathcal{A}_{\it rz} = -\frac{1}{De}\mathcal{A}_{\it rz},
\label{Arz_ND long}
\end{equation}
\begin{equation}
 U_{r} \frac{\partial \mathcal{A}_{\it rr}}{\partial R}-2\frac{\partial U_{r}}{\partial Z} \mathcal{A}_{\it rz}-2\frac{\partial U_{r}}{\partial R} \mathcal{A}_{\it rr} = -\frac{1}{De}(\mathcal{A}_{\it rr} -\epsilon^2),
\label{Arr_ND long}
\end{equation}
where the viscosity ratio $\tilde{\beta}$ and the Deborah number $De$ are defined in (\ref{beta and beta_t}) and (\ref{De and Wi}), respectively. 
Finally, considering the leading order in $\epsilon$, we obtain the non-dimensional governing equations (\ref{ND_gov}).

\section{Details of finite-element numerical simulations}\label{App B}

\begin{table}
  \begin{center}
\def~{\hphantom{0}}
  \begin{tabular}{cccccccccccc}
   $r_{\it in}$ &   $r_{\it out}$ & $h$& $\eta_0$ & $\rho$  & $q$  & $\lambda$  & $De$ & $Wi $&$\tilde{\beta}$ &$\alpha$&$\epsilon$\\
     (mm)  &  (mm) & (mm) & (Pa s) & (kg m$^{-3}$) & (m${^3}$ s$^{-1}$)  & (s)& (--)& (--)  &(--)&(--) & (--)\\
    %\hline
     10 & 100 &4 & 1 & 1 & $5.03\times10^{-4}$ & $0-0.23$ &$0-0.46$& $0-11.5$& $0.05-0.5$ & 0.1& 0.04
    \end{tabular}
  \caption{Values of physical and geometrical parameters used in the axisymmetric finite-element numerical simulations of the radial flow of the Oldroyd-B fluid between two parallel plates. The relaxation time $\lambda$ is adjusted to obtain the desired value of the Deborah number $De$ according to (\ref{De and Wi}). The Reynolds number is given by $Re=\rho u_c h/\eta_0=8\times10^{-4}$ and the characteristic pressure $p_c$ is given by $p_c=\eta_0u_cr_{\it out}/h^2=1250$ Pa, where $u_c= q/(2\pi r_{\it out}h)=0.2$ m s$^{-1}$ is the characteristic velocity.}
  \label{T1}
  \end{center}
\end{table}

In this appendix, we describe the numerical procedure used to solve the system of nonlinear governing equations (\ref{Continuity+Momentum})--(\ref{A}) for the radial flow of the Oldroyd-B fluid.
We have performed finite-element numerical simulations using the viscoelastic flow module in COMSOL Multiphysics, which includes the Oldroyd-B constitutive model (version 6.2, COMSOL AB, Stockholm, Sweden). 

We exploit the axial symmetry of the problem to reduce the three-dimensional configuration to a two-dimensional formulation and solve only the cross-section of the geometry.
 We impose the no-slip and no-penetration boundary conditions along the walls and fully developed unidirectional flow with the flow rate $q$ at the entrance of the narrow tube of radius $r_{\it in}$. At the outlet, $r=r_{\it out}$, the reference value for the pressure is set to zero on $z=0$.
 Finally, we calculate the pressure drop between a given location ($r=\alpha_i r_{\it out}$) and the outlet ($r=r_{\it out}$) at the midplane $z=0$, defined as $\Delta p=p(r=\alpha_i r_{\it out},z=0)-p(r=r_{\it out},z=0)$.

We summarize in table~\ref{T1} the values of physical and geometrical parameters used in the numerical simulations.
We consider a configuration with an aspect ratio $\epsilon=h/r_{\it out}=0.04$ and the inlet-to-outlet aspect ratio $\alpha=r_{\it in}/r_{\it out}=0.1$, and study four different polymer-to-total viscosity ratios: $\tilde{\beta}=\eta_p/\eta_0=
0.05,$ $0.1$, $0.2$, and $0.5$, where the first value corresponds to the ultra-dilute limit. 
To facilitate the simulations, we round the corner between the inlet tube and the top plate to a quarter-circle of radius $r_{\it cor}=h=4$ mm, and introduce a dimensionless parameter $\alpha_{\it cor}=(r_{\it in}+r_{\it cor})/r_{\it out}=0.14$. The rounded geometry improves convergence compared with the unrounded configuration and enables us to obtain converged solutions at higher $De$.

To study the effect of different Deborah numbers in the ultra-dilute limit, $\tilde{\beta}=0.05$, we vary the value of the relaxation time $\lambda$ from $0$ s to $0.23$ s to change $De$ from $0$ to $0.46$, while keeping the values of all other parameters. 
Similarly, to examine the effect of the polymer-to-total viscosity ratio $\tilde{\beta}=\eta_p/\eta_0$ on the pressure drop, we vary $\eta_s$ and $\eta_p$ while fixing $\eta_0=1$ Pa s. For each value of $\tilde{\beta}$, we perform simulations over a range of $\lambda$, keeping all other parameters fixed. 
In steady-state COMSOL simulations, we could not obtain converged results beyond $De=0.46$ ($Wi=11.5$) for $\tilde{\beta}=0.05$. As we increase $\tilde{\beta}$, the maximum Deborah number for which we obtain converged solutions decreases (see figure~\ref{F8}). Specifically, the maximum value is $De=0.28$ ($Wi=7$) for $\tilde{\beta}=0.1$, $De=0.2$ ($Wi=5$) for $\tilde{\beta}=0.2$, and $De=0.12$ ($Wi=3$) for $\tilde{\beta}=0.5$. Fluid inertia is negligible in our finite-element simulations, as the reduced Reynolds number, $\epsilon Re=(h/r_{\it out})\rho u_ch/\eta_0=3.2\times10^{-5}$, is extremely small.

We discretize the velocity field with second-order Lagrange elements and the pressure and polymer stress fields with first-order Lagrange elements, yielding a mesh of approximately $7\times10^{3}$ elements, and solve the resulting system using the PARDISO solver in COMSOL Multiphysics with a relative tolerance of $10^{-5}$. We have performed grid sensitivity tests and established grid independence at this resolution. In particular, for $\tilde{\beta}=0.05$, we carried out simulations using two different meshes containing approximately $7\times10^{3}$ and $3.6\times10^{4}$ elements. We varied the Deborah number from 0 to 0.2 in increments of 0.002 and found a maximum relative error of 0.6 \% in the pressure drop.

\bibliographystyle{jfm}
\bibliography{literature}

%\bibliographystyle{jfm}
%\bibliography{jfm2esam}
\end{document}